\documentclass[
prb,
aps,
amssymb,
twocolumn,
superscriptaddress]{revtex4-2}

\usepackage[english]{babel}
\usepackage[utf8]{inputenc}
\usepackage{graphicx}
\usepackage{amsmath}
\usepackage{systeme}
\usepackage{bbold}
\usepackage{amsthm}
\usepackage{dsfont}
\usepackage{mathtools}
\usepackage{physics}
\usepackage{soul}
\usepackage{ulem}
\usepackage{xcolor}
\usepackage{multirow}
\usepackage{adjustbox}
\usepackage{placeins}
\usepackage[T1]{fontenc}
\usepackage{lipsum}
\usepackage{csquotes}
\usepackage{hyperref}
\hypersetup{colorlinks=true}

\usepackage{rotating}
\definecolor{pink}{rgb}{1.0, 0.33, 0.64}

\begin{document}

\title{Transversality-Enforced Tight-Binding Models for 3D Photonic Crystals aided by Topological Quantum Chemistry}

\author{Antonio Morales-P\'erez*}
\email{antonio.morales@dipc.org}
\affiliation{Donostia International Physics Center, Paseo Manuel de Lardizabal 4, 20018 Donostia-San Sebastian, Spain.}
\affiliation{Material and Applied Physics Department, University of the Basque Country (UPV/EHU), Donostia-San Sebastián, Spain.}

\author{Chiara Devescovi*}
\affiliation{Donostia International Physics Center, Paseo Manuel de Lardizabal 4, 20018 Donostia-San Sebastian, Spain.}

\author{Yoonseok Hwang}
\affiliation{Department of Physics, University of Illinois at Urbana-Champaign, Urbana, IL, USA}

\author{Mikel García-Díez}
\affiliation{Donostia International Physics Center, Paseo Manuel de Lardizabal 4, 20018 Donostia-San Sebastian, Spain.}
\affiliation{Physics Department, University of the Basque Country (UPV/EHU), Bilbao, Spain}

\author{Barry Bradlyn}
\affiliation{Department of Physics, University of Illinois at Urbana-Champaign, Urbana, IL, USA}

\author{Juan L. Ma\~nes}
\affiliation{Physics Department, University of the Basque Country (UPV/EHU), Bilbao, Spain}

\author{Maia G. Vergniory}
\email{maia.vergniory@cpfs.mpg.de}
\affiliation{Donostia International Physics Center, Paseo Manuel de Lardizabal 4, 20018 Donostia-San Sebastian, Spain.}
\affiliation{Département de Physique et Institut Quantique, Université de Sherbrooke, Sherbrooke, QC J1K 2R1 Canada.}

\author{Aitzol Garc\'ia-Etxarri}
\email{aitzolgarcia@dipc.org}
\affiliation{Donostia International Physics Center, Paseo Manuel de Lardizabal 4, 20018 Donostia-San Sebastian, Spain.}
\affiliation{IKERBASQUE, Basque Foundation for Science, Mar\'ia D\'iaz de Haro 3, 48013 Bilbao, Spain.}

\date{\today}

\begin{abstract}

Tight-binding models can accurately replicate the band structure and topology of crystalline systems. They have been widely used in solid-state physics due to their versatility and low computational cost.
It is straightforward to build an accurate tight-binding model of any crystalline system using the crystal's maximally localized Wannier functions as a basis.
Unfortunately, in 3D photonic crystals, the transversality condition of Maxwell's equations precludes the construction of a basis of maximally localized Wannier functions via usual techniques. As a result, building reliable tight-binding models of 3D Photonic Crystals has not been straightforward up to now.
In this work, we show how to overcome this problem using topological quantum chemistry, allowing us to express the band structure of the photonic crystal as a difference of band representations.
This can be achieved by introducing a set of auxiliary modes, as recently proposed in \textit{Christensen et al.}, Phys. Rev. X {\bf 12}, 021066 (2022), which regularizes the $\Gamma$-point obstruction arising from the transversality constraint of Maxwell's equations.
The decomposition into elementary band representations allows us to isolate a set of pseudo-orbitals that permit us to construct an accurate transversality-enforced tight-binding model that matches the dispersion, symmetry content, and topology of the 3D photonic crystal under study. 
Moreover, we show how to introduce the effects of a gyrotropic bias in the framework, modeled via non-minimal coupling to a static magnetic field.
Our work provides the first systematic method to analytically model the photonic bands of the lowest transverse modes over the entire Brillouin zone via a transversality-enforced tight-binding model.

\end{abstract}

\maketitle

\section{Introduction} \label{sec: intro}

Photonic Crystals (PhCs) were first conceived in the 1980s~\cite{sajeev1987photoniccrystal} as optical systems with periodic refractive indices.
Just as in solid-state materials, this periodicity can give rise to band gaps where light in a given frequency range is blocked from propagating through the structure.
Since then, researchers have used PhCs to guide and control light propagation in various technological devices, such as fiber optics~\cite{russell2003photonic}.
Moreover, the understanding of the propagation of light in periodic media has also allowed us to understand many optical phenomena in nature, for example, the iridescent colors of the wings of certain butterfly species \cite{PhCreview2001Yablonovitch,peacock2003Zi,butterfly2011Yu,teyssier2015photonic,wilts2012brilliant,parker2001aphrodite}.

Recently, the field has experienced a rebirth due to the discovery of the topological properties that PhCs may sustain.
Topological PhCs, when interfaced with trivial dielectric materials, can sustain surface and edge states within the bulk band gap.
Importantly, these states are topologically protected and constrain light propagation on an interface, allowing for an enhanced local density of states capable of boosting light-matter interactions in confined regions. 
Contemporary theoretical studies envision novel physical phenomena and topological phases in these systems \cite{devescovi2021cubic,topoPhotoRev}.
Nevertheless, the development of such intricate studies requires the simulation of very large and complicated supercells, which often reach the limits of state-of-the-art high-performance computing resources 
\footnote{For instance, the slab-geometry supercell calculations performed in 3D topological PhCs of Ref. \cite{devescovi2021cubic} required 300 Gb of RAM and several weeks of runtime per each k-point. Scaling these calculations to rod- and cube-geometry supercells would be prohibitively expensive.}.
Therefore, developing highly efficient simulation techniques is in high demand for the field's growth.

The tight-binding (TB) approach, widely used in solid-state physics, can reproduce a crystal's band structure and topological properties while using a small set of parameters and basis functions compared to \textit{ab initio} models or exact electromagnetic solvers.
Thus, these models allow the simulation of much larger systems using fewer computational resources. 

Unfortunately, extending the TB approach from electronics to PhCs is not straightforward.
Electronic systems naturally display a set of orbitals that can be mathematically mapped to a set of exponentially confined states known as maximally localized Wannier Functions (WFs).
This set of functions is, ideally, exponentially localized and thus well-suited for a basis for a TB model. As we will see, 
finding such a basis for confined states in PhCs is not trivial.

In 1D and 2D PhCs, decoupling the solutions to the wave equation into scalar transverse electric (TE) and transverse magnetic (TM) modes is always possible.
By doing so, one can express Maxwell's equations as a scalar eigenproblem directly analogous to the spinless Schrödinger equation.
In such scenarios, a photonic equivalent of maximally-localized WFs can always be found \cite{gupta2022wannier, romano2018wannier, albert2000generalized, busch2003wannier} following standard solid-state Wannierization methods.
Consequently, one can use these functions directly as a basis set of states to construct a reliable TB model of the crystal under study.

Unfortunately, the vectorial nature of the electromagnetic field cannot be avoided in the theoretical description of 3D PhCs, making it impossible to find maximally localized WFs to use as a basis for the TB model.
The transverse nature of the electromagnetic solutions makes it impossible to properly define 3D electromagnetic plane waves in reciprocal space at frequency $\omega=0$ and wavevector $k=0$ (the $\Gamma$ point).
In 2D, any TE or TM mode with constant amplitude and polarization vector $\hat{u}$ pointing in the $\hat{z}$ direction is a valid solution at the $\Gamma$ point at zero frequency.
This description in $\mathbf{k}$-space is valid because $\hat{u}=\hat{z}$ is orthogonal to $k_x$ and $k_y$, respecting the transversality condition.
In 3D, the zero frequency constraint also forces the solution to be of constant amplitude, but no $\hat{u}$ will be able to fulfill the transversality condition in every direction.
This leads to a singularity at $\Gamma$ in the eigenvectors of any band structure of 3D PhCs at zero frequency.
This singularity is not an obstacle to analyzing the band structure, identifying band gaps, or even determining the topological character of a photonic crystal through, for example, Wilson loop calculations \cite{devescovi2024tutorial}.
Nevertheless, it impedes the application of topological quantum chemistry (TQC) techniques to the first set of bands \cite{bradlyn2017topological}, and importantly for us, also obstructs the construction of maximally localized photonic WFs, as proved in Refs. \cite{wolff2013generation, busch2003wannier,christensen2022location}.

Mathematically, the $\Gamma$ point singularity at $\omega=0$ originates from the divergence-free condition of the Maxwell equation, $\grad\cdot\textbf{B}=0$. 
In a PhC, this condition translates in reciprocal space into

\begin{equation}
\label{divergence}
    (\textbf{k}+\textbf{G})\cdot\textbf{B}^{(\textbf{G})}_{n,\textbf{k}}=0,
\end{equation}
where $\textbf{k}$ is a vector in the first Brillouin zone (BZ),
$\textbf{G}$ denotes a reciprocal lattice vector and 
$\textbf{B}^{(\textbf{G})}_{n,\textbf{k}}$ are the Fourier components of the Bloch magnetic field, with $n$ the band index.
Geometrically, Eq.~\eqref{divergence} restricts field solutions to lie in the tangent plane of a 3D sphere with radius $|\textbf{k}+\textbf{G}|$ centered around the Gamma ($\Gamma$) point.
If we take the limit of $\mathbf{k}\rightarrow 0$, then, for the lowest bands of the photonic crystal, only the $\mathbf{G}=0$ components of $\mathbf{B}^{(\mathbf{G})}_{n\mathbf{k}}$ are nonzero to leading order for the lowest two bands \cite{datta1993effective,krokhin2002long}. 
This means that for the lowest bands near the $\Gamma$ point, the Bloch fields are constrained to be tangent to a small sphere in momentum space surrounding $\Gamma$.
Notably, since the Euler characteristic of (the tangent space of) the 3D sphere is nonzero \cite{milnor1978analytic,eisenberg1979proof}, a vector field in the tangent space of the sphere will always be singular at one point.
This lack of analyticity in the Bloch basis precludes the construction of exponentially localized WFs \cite{wolff2013generation} for the lowest photonic bands.
In simple terms, this is because the WFs are the Fourier transform of the Bloch functions, which are discontinuous in this case.
Furthermore, Eq.~\eqref{divergence} indicates that polarization is indeterminate for the $\omega=0$ modes at $\Gamma$.

In this article, we propose a novel method for constructing a reliable TB representation of 3D PhCs, overcoming the absence of maximally localized WFs in 3D. To do so, we use TQC, which will allow us to express the band structure as a sum of elementary band representations (EBRs). The bases of those EBRs can be used as localized orbitals to build a TB model without the need to obtain maximally localized WFs.

We achieve this by adding auxiliary longitudinal modes to the physical transverse bands of the PhC, able to regularize the $\Gamma$ point singularity \cite{christensen2022location}, which arises from the transversality constraint of the Maxwell equations.
This allows us to construct a Transversality-Enforced Tight-Binding model (TETB) that accurately captures the photonic bands' energy dispersion, symmetry, and topology.
We will focus our study on source-free PhCs made of non-frequency dispersive and non-bianisotropic materials.
The method proposed is based solely on group theoretical arguments and exploits the formalism of TQC for non-fermionic systems \cite{JL20phonontopo,martin23phononApp,xu2022catalogue}.

As a first step, we use TQC to express the band structure as a sum of elementary band representations (EBRs), which we can adopt as a starting basis for constructing a photonic TETB.
As shown in section II, to achieve this, we need to solve a Diophantine system of equations.
We approach this problem using three complementary approaches:
\begin{enumerate}
    \item an \textit{optimization method} that takes the number of bands as an objective function to minimize,
    \item an \textit{enumeration method} that searches for all possible solutions with a given number of bands and
    \item a different approach using a \textit{real-space invariants (RSI)}~\cite{song2020twisted,xu2021filling} \textit{based method} to define an auxiliary problem that, in certain situations, can be easier to solve than the original Diophantine problem.
\end{enumerate}

These three methods are alternative approaches to solving the same problem. We include them to demonstrate how different methods can lead to the same outcome in some cases, even though some may be more effective than others. Specifically, the optimization method will provide, at best, one solution with an optimal (minimal) number of bands.
As the method is based on an optimization algorithm, it may get stuck in a local minimal solution which is not a global minimum.
Additionally, as explained later,  the method is blind to a whole class of optimal solutions. 
On the other hand, the enumeration method will address those problems by separating the initial integer problem into two non-negative integer problems, making it much easier to solve. This approach will always provide all the optimal solutions. Lastly, the RSI-based method relies on defining an auxiliary problem, which in certain SGs could be easier to solve but might be even more complicated than the original problem in other cases. In all cases, the obtained sum of EBRs defines a set of pseudo-orbitals and their corresponding real space locations. 
As a concrete example, we consider the crystal shown in Fig.~\ref{fig:panel224}\textcolor{red}{a}, whose space group (SG) is $Pn\Bar{3}m$ (No. 224). This example has been thoroughly studied by our group \cite{devescovi2021cubic,devescovi2023axion}, so it is a perfect test for our new method. Additionally, it is an interesting example due to the triple degeneracy that occurs at the $R$-point in $k$-space isolated in energy and protected by symmetry. This degeneracy can be broken by the presence of a magnetic field, forming Weyl points isolated in energy. This is an example of a photonic Weyl semimetal whose Weyl's point can be moved along the BZ by modulating the magnetic field applied.

Next, we show how to use the set of pseudo-orbitals to build a TETB model replicating the band structure of the 3D PhC.
Comparing it to the bands obtained via the numerical software MIT Photonic Bands (MPB)~\cite{johnson2001block}, we verify that the TETB constitutes a reliable crystal model, replicating the dispersion properties, symmetry content, and topology of the PhC.
Additionally, we explain how to introduce the effects of a static magnetic field into the model and check the validity of the proposed method by comparison with MPB calculations. 

Our conclusions and outlook are presented in section III. At the same time, technical details and explicit step-by-step examples of the application of the three methods for solving the Diophantine system of equations can be found in the Supplementary Material (SM) \cite{SuppMat} (see also Refs. \cite{cano2021band,hwang2019fragile,van2018higher,perez2015symmetry,watanabe2018space} therein), where we also consider another example of a crystal with SG $Pm\Bar{3}m$ (No. 221). This is a simpler case compared to the crystal with SG $Pn\Bar{3}m$ (No. 224) and it is a perfect system to introduce a more didactical deduction of the TETB model.
Additionally, since the enumeration method is the only one that will give all optimal solutions in all cases, we have developed a code that implements such method and posted it on GitHub~\cite{github}. The functions necessary to implement the algorithm are already predefined in the provided GitHub repository, making the method user-friendly. Moreover, some examples are provided to help understand how the code works.

\section{Methods} \label{sec: meth}

As previously stated, our first objective is to identify a basis of pseudo-orbitals that can be used to build a TETB model for a 3D PhC.
To do this, first, we need to pinpoint the symmetry content of the band structure at all high symmetry points (HSPs) in the BZ.
This is determined by the irreducible representations (irreps) of the eigenvectors at the HSPs.
These irreps constitute the first stage of studying the system's topology via TQC.
Any reliable TETB model of a crystal, which replicates the system's topology, must reproduce such irreps at the HSPs.

The irreps of the electromagnetic fields at the HSPs are computed numerically.
The procedure has to be carried out on a bundle of bands that can form an isolated set that satisfies the compatibility relations and has trivial symmetry indicators.
The size of this bundle depends on the number of bands one wants to include in the TETB description of the PhC. 
We used the Julia package MPBUtils.jl \cite{MPBUtils} at the time of writing (v0.1.11) dedicated to the computation of the symmetry eigenvalues of any mode from overlap integrals and, later, we applied Schur's orthogonality relations~\cite{miller1973symmetry} to obtain the irrep decomposition.

We label the irreps following the notation of the Bilbao Crystallographic Server (BCS)~\cite{elcoro2017double,aroyo2011crystallography}.
We denote each irrep as $\rho_{i[\textbf{k}]}$ where $i$ labels the particular irrep at each class of HSP $[\textbf{k}]$.
As mentioned above, in 3D PhCs, the irrep content can be computed at every HSP except for the zero-frequency modes at $\Gamma$, where the symmetry identification is ill-defined.
In the following, the symbol $(\blacksquare)^{2T}$ will represent this ill-defined irrep content or ``surrogate representation'', consistent with the notation of Ref.~\cite{christensen2022location}.
Thus, we write the complete ``transverse symmetry vector'' ($\textbf{v}^T$) as

\begin{equation}
  \textbf{v}^T= \bigoplus_{i,[\textbf{k}]} n_{i[\textbf{k}]}\rho_{i[\textbf{k}]} \oplus (\blacksquare)^{2T} \equiv \tilde{\textbf{v}}^T\oplus (\blacksquare)^{2T},
\label{complete}
\end{equation}
where $n_{i[\textbf{k}]}$ labels the multiplicity of each irrep and $\tilde{\textbf{v}}^T$ excludes the zero frequency modes at $\Gamma$.

Using TQC, we aim to decompose $\textbf{v}^T$ into elementary band representations (EBRs), which will constitute the building blocks of our TETB model.
Roughly speaking, an EBR describes the transformation properties of the bands induced by a set of orbitals. Those orbitals transform under a certain irrep $\rho_\mathbf{q}$ of the site-symmetry group and are placed at a particular maximal Wyckoff Position (WP) $\mathbf{q}$.
Thus, a certain EBR can be specified by the pair `$\rho_\mathbf{q}@\mathbf{q}$'.
Note that even if this description is accurate in solid-state physics, the concept of orbital is not technically precise in PhCs.
Therefore, we use `pseudo-orbitals' to refer to the localized, symmetric functions that will induce the photonic bands.
All EBRs for all SGs and the irreps at the HSPs they induce are tabulated in the BCS.
Then, the initial problem reduces to finding a linear combination of EBRs, $\sum_a n_a \text{EBR}_a$ with $n_a \in \mathds{Z}$, that induces the symmetry content of $\textbf{v}^T$. The following paragraph shows that this problem can be expressed as a linear system of equations with integer coefficients.

First, we collect all the EBRs of the SG of the crystal into a matrix $\textbf{A}$.
Denoting the number of EBRs in the SG by $N_{EBR}$ and the total number irreps at all HSPs by $N_{irr}$, $\textbf{A}$ has a size of $N_{irr}\times N_{EBR}$.
The $i$-th column of $\textbf{A}$ will represent the symmetry vector of the $i$-th EBR, as tabulated in the BCS.
Then, to find a plausible linear combination of EBRs that describes the symmetry content of $\textbf{v}^T$, we must obtain an integer solution to

\begin{equation}
    \textbf{v}^T=\textbf{A}\textbf{n}^T,
\label{matrix_eq}
\end{equation}
where $\textbf{n}^T$ is an integer vector describing the multiplicities $n_a$ of the $N_{EBR}$ EBRs. An example showing the construction of these matrices and vectors can be found in the SM \cite{SuppMat} (see also Refs. \cite{cano2021band,hwang2019fragile,van2018higher,perez2015symmetry,watanabe2018space} therein).

It is worth noticing that since $\mathbf{v}^T$ represents an isolated trivial set of bands, $\textbf{n}^T$ will always be an integer vector.
In general, $\text{rank}(\textbf{A})\leq\min(N_{EBR}, N_{irr})$, which means that we can, in principle, find infinite solutions to Eq.~\eqref{matrix_eq}.
Such problems are known as indeterminate Diophantine equations.
From a physical perspective, the existence of multiple solutions is attributable to the fact that different linear combinations of EBRs can yield the same symmetry vector, resulting in an equivalent description of the photonic band structure.
From a mathematical viewpoint, the existence of infinitely many solutions is related to the non-positivity of the components of $\mathbf{n}^T$, which can be arbitrarily large while keeping the components of $\mathbf{v}^T$ finite and fixed.


Note that the standard methods to solve linear systems of equations are not well suited for solving Eq.~\eqref{matrix_eq} since they yield real-valued solutions while we need integer solutions. 
A convenient way to parametrize the solutions to this equation is to compute the Smith decomposition of the integer matrix $\textbf{A}$

\begin{equation}
\label{smithA}
    \textbf{A}=\textbf{U}^{-1}\textbf{D}\textbf{V}^{-1},
\end{equation}
where $\textbf{U}$ and $\textbf{V}$ are matrices invertible over the integers
and $\textbf{D}$ is an integer diagonal matrix.
Then, the general solution $\textbf{n}^T$ can be written as 

\begin{equation}
    \textbf{n}^T = \textbf{V} \textbf{D}^{+} \textbf{U} \textbf{v}^T + (\mathds{1}_{N_{EBR}} - \textbf{V} \textbf{D}^{+} \textbf{U} \textbf{A}) \textbf{z}
\label{pseudoinverse}
\end{equation}
where $\textbf{D}^+$ denotes the pseudo-inverse of $\mathbf{D}$, 
$\textbf{z}$ is an arbitrary integer vector, and the pseudo-inverse of a diagonal matrix is obtained by 
replacing each non-zero eigenvalue by its inverse $\lambda_i \to 1/\lambda_i$. See the SM \cite{SuppMat} (see also Refs. \cite{cano2021band,hwang2019fragile,van2018higher,perez2015symmetry,watanabe2018space} therein) for a derivation of this equation.

Eq.~\eqref{pseudoinverse} provides an infinite number of solutions, one for each choice of the integer vector $\textbf{z}$.
If all the multiplicities $n_a$ in a given solution were positive, then for each \hbox{$\text{EBR}_a = \rho_\mathbf{q}@\mathbf{q}$} the corresponding  TETB model would include $n_a$ orbitals transforming according to the irrep $\rho_\mathbf{q}$ at the maximal Wyckoff Position $\mathbf{q}$.  However, most solutions to Eq.~\eqref{matrix_eq} involve both positive and negative EBR multiplicities. The appearance of negative coefficients while trying to express the lowest bands of a 3D PhC as a linear combination of EBRs is reminiscent of the concept of `fragility' in topological band theory \cite{po2018fragile,de2019engineering,blanco2020tutorial}.
This seeming fragility can be exploited in the development of the TETB model by reinterpreting the EBRs with negative coefficients as a set of `auxiliary' longitudinal modes ($\mathbf{v}^L$) which regularize the $\Gamma$ point surrogate representation (see Sec. \ref{subsec: mapping} for more details).
This reorganization of the modes can be formalized by expressing the vector $\mathbf{n}^T$ as the difference of two positive vectors $\textbf{n}^T=\textbf{n}^{T+L}-\textbf{n}^L$. Then, moving the negative EBR multiplicities to the left-hand side,  Eq.~\eqref{matrix_eq} can be written
\begin{equation}
\textbf{v}^T+\textbf{A}\textbf{n}^L=\textbf{A}\textbf{n}^{T+L}
\label{LandT}
\end{equation}
Thus, the EBRs with positive multiplicities in $\textbf{n}^T$ determine the pseudo-orbitals to be included in the TETB model, while those with negative multiplicities describe the auxiliary bands used to regularize the $\Gamma$ point singularity \cite{christensen2022location}.  
In terms of  $\textbf{v}^L=\textbf{A}\textbf{n}^L$ and $\textbf{v}^{T+L}=\textbf{A} \textbf{n}^{T+L}$, Eq. (\ref{LandT}) can be rewritten as 
\begin{equation}
\textbf{v}^T+\textbf{v}^L=\textbf{v}^{T+L}.
\label{vLandT}
\end{equation}
Thus, unless all EBR multiplicities in $\mathbf{n}^T$ are positive, the TETB model will describe not just the physical transverse bands but also a set of auxiliary bands with symmetry content defined by $\textbf{v}^L=\textbf{A}\textbf{n}^L$.
Note that the number of transverse bands $\mu^T$ is fixed \textit{a priori} by our choice of the symmetry vector $\mathbf{v}^T$, but the number of auxiliary bands $\mu^L$ will be solution dependent.  To simplify the construction of the TETB model, we would like to minimize the number of auxiliary bands. This motivates the introduction of a norm in the space of solutions defined by $||\textbf{n}|| = \sum_{a=1}^{N_{EBR}} d_a  |n_a|$, where $d_a$ is the dimension of the  EBR and $|n_a|$ denotes the absolute value of its multiplicity $n_a$. Noting that $||\textbf{n}^T||=\mu^{T+L}+\mu^L=\mu^T + 2 \mu^L$, it is apparent that minimizing $||\textbf{n}^T||$ also minimizes the number of auxiliary bands and the complexity of the TETB model.
Although the easiest way to solve Eq.~\eqref{matrix_eq} is to set $\mathbf{z}=0$ in Eq.~\eqref{pseudoinverse} and
this provides a solution; it is not guaranteed to have a minimal norm, and, in general, the resulting TETB model will include more auxiliary bands than is strictly necessary.

\subsection{Solving the Diophantine equations} \label{subsec: solving}

Before discussing the three strategies proposed in this paper to solve Eq.~\eqref{matrix_eq}, we must further address the problem of the zero-energy singularity at $\Gamma$. Namely, we should discuss how to assign a concrete set of irreps (the surrogate representation) to $(\blacksquare)^{2T}$ in Eq. \eqref{complete}. This can be done by imposing compatibility relations (continuity) from symmetry lines and planes intersecting at $\Gamma$. For the 113 space groups without inversion or roto-inversions, the symmetry content at $\Gamma$ is uniquely defined or \textit{pinned} by the compatibility relations \cite{christensen2022location}. For the remaining 117 space groups, the symmetry content at $\Gamma$ is only partially constrained by the compatibility relations, and infinitely many assignments of irreps are possible. These possibilities have been parametrized and tabulated in Ref. \cite{christensen2022location} where, for each of the \textit{unpinned} groups, a particular solution together with a basis spanning the space of solutions to the compatibility relations on the surrogate representation are given. Although one could be tempted to use the tabulated particular solution, this is not always the best choice, as shown in the examples in the SM \cite{SuppMat} (see also Refs. \cite{cano2021band,hwang2019fragile,van2018higher,perez2015symmetry,watanabe2018space} therein). The reason is that a solution's minimal number of auxiliary bands strongly depends on the specific set of irreps in the surrogate representation $(\blacksquare)^{2T}$. Thus, it would seem that before we can even attempt to find an optimal solution to the Diophantine problem, we must be lucky enough to choose the appropriate symmetry content at $\Gamma$. 

The way out of this conundrum is to  define a new vector $\textbf{v}^T_{(BZ-\Gamma)}$ by eliminating the symmetry content at $\Gamma$

\begin{equation}
\textbf{v}^T_{(BZ-\Gamma)} = \bigoplus n_{i[\textbf{k}\neq 0]} \rho_{i[\textbf{k}\neq 0]}.
\label{defective}
\end{equation}
To match the dimensionality of Eq.~\eqref{defective}, we eliminate the irreps at the $\Gamma$ point from the EBR matrix $\textbf{A}$, resulting in a new matrix $\textbf{A}_{(BZ-\Gamma)}$ of size $(N_{irr}-N_{irr[\Gamma]})\times N_{EBR}$, where $N_{irr[\Gamma]}$ is the number of irreps at $\Gamma$. This makes the approach entirely agnostic as to the singular content at $\Gamma$, which, as shown below, becomes an output of the procedure. The resulting reduced Diophantine problem will be

\begin{equation}
\textbf{v}^T_{(BZ-\Gamma)}=\textbf{A}_{(BZ-\Gamma)}\textbf{n}^T
\label{diophantine}
\end{equation}
where $\textbf{n}^T=\textbf{n}^{T+L}-\textbf{n}^L$ specifies both the pseudo-orbitals to be used in the TETB 
(through $\textbf{n}^{T+L}$) and the symmetry content of the auxiliary bands (through $\textbf{n}^L$). The surrogate representation can be uniquely determined by combining  Eqs. (\ref{complete}) and (\ref{vLandT}),

\begin{equation}
(\blacksquare)^{2T}= \textbf{v}^{T+L}-\textbf{v}^L-\tilde{\textbf{v}}^T,
\label{BS} 
\end{equation}
where $\textbf{v}^L=\textbf{A}\textbf{n}^L$, $\textbf{v}^{T+L}=\textbf{A}\textbf{n}^{T+L}$ and $\tilde{\textbf{v}}^T$ has been defined in Eq.~\eqref{complete}. 
Note that, by construction, every solution to Eq. (\ref{diophantine}) is automatically a solution to Eq. (\ref{matrix_eq}) for \textit{that} particular singular content at $\Gamma$. Conversely,  given that Eq. (\ref{diophantine}) does not use the information at $\Gamma$ and is, therefore, less restrictive, it follows that every solution to Eq. (\ref{matrix_eq}) must also be a solution to  Eq. (\ref{diophantine}). Thus, solving Eq. (\ref{diophantine}) is equivalent to solving Eq. (\ref{matrix_eq}) for \textit{all} possible choices of the surrogate representation. This guarantees that, in principle, an optimal TETB model can always be found.

We are now ready to discuss the strategies enumerated in the introduction to find optimal solutions to the Diophantine equations i.e., solutions with the smallest possible number of auxiliary bands.  In the \textit{optimization method}, the norm \hbox{$||\textbf{n}|| = \sum_{a=1}^{N_{EBR}} d_a  |n_a|$} is taken as the objective function to minimize,  subject to the constraints in Eq. \eqref{diophantine}. Thus, the surrogate representation is \textit{not} an input in this method but part of its output, which takes the form $({\textbf{v}}^T,(\blacksquare)^{2T})$. Note that, as mentioned above, the surrogate representation that minimizes the number of auxiliary bands might differ from the one tabulated in Ref. \cite{christensen2022location}. 

A common issue with optimization routines is that they are not guaranteed to find the absolute minimum and can easily get stuck around local minima. Indeed, unless the solution contains the minimum number of auxiliary bands allowed by the EBR dimensions for the particular SG, it is very difficult to ascertain that there are no solutions with fewer auxiliary bands. 

A second, more fundamental limitation of the optimization method is the somewhat counterintuitive fact that the correspondence between solutions to the Diophantine equations and TETB models is not one-to-one in general. On the one hand, there are solutions to Eqs. \eqref{matrix_eq} or \eqref{diophantine}, which can not be used to build a viable TETB model. In those \textit{unphysical} solutions, there is a cancellation between an irrep in the transverse spectrum and a negative multiplicity irrep in the singular content at $\Gamma$, with the result that the TETB displays \textit{three} rather than two transverse bands with zero energy at  $\Gamma$, as shown in the SM \cite{SuppMat} (see also Refs. \cite{cano2021band,hwang2019fragile,van2018higher,perez2015symmetry,watanabe2018space} therein). On the other hand, there are many cases where the optimal TETB model for a particular crystal has at least one EBR, which is common to $\textbf{n}^{T+L}$ and $\textbf{n}^L$. Obviously, the common EBR will always cancel out in the vector $\textbf{n}^T=\textbf{n}^{T+L}-\textbf{n}^L$ and will be invisible among the solutions of the Diophantine equations. For lack of a better name, we will refer to these models as \textit{composite} TETB models. See the SM \cite{SuppMat} (see also Refs. \cite{cano2021band,hwang2019fragile,van2018higher,perez2015symmetry,watanabe2018space} therein) for examples. 
 
While unphysical solutions can be easily screened out as explained in the SM \cite{SuppMat} (see also Refs. \cite{cano2021band,hwang2019fragile,van2018higher,perez2015symmetry,watanabe2018space} therein) and do not pose a problem for the optimization method, the possibility of optimal but composite TETB models requires a recasting of the Diophantine equations and was the main motivation behind the development of the \textit{enumeration method}. As mentioned above, the non-positivity of $\mathbf{n}^T$ is behind the existence of infinitely many solutions, which poses a practical problem. The key idea is to recast Eq. \eqref{diophantine}, where the components of the solution vector $\mathbf{n}^T$ can be positive or negative integers in a manner where all components are positive. Using $\mathbf{n}^T=\mathbf{n}^{T+L}-\mathbf{n}^L$ and moving $\mathbf{n}^L$ to the left hand side, Eq. \eqref{diophantine} becomes

\begin{equation}
    \textbf{v}^T_{(BZ-\Gamma)}+\textbf{A}_{(BZ-\Gamma)}\textbf{n}^L =\textbf{A}_{(BZ-\Gamma)}\textbf{n}^{T+L},
\label{enum}
\end{equation}
which is solved in two steps. First, we find all the positive vectors $\textbf{n}^L$ satisfying $||\textbf{n}^L||=\mu^L$, i.e., we enumerate \textit{all} possible sets of $\mu^L$ auxiliary bands. In the second step, we plug each of the obtained $\textbf{n}^L$ into the left-hand side of Eq. \eqref{enum} and solve for $\textbf{n}^{T+L}$. These two problems involve only positive integer vectors with a finite number of solutions that can be found very efficiently.
Note that the output of this method takes the form  $(\textbf{n}^L,\textbf{n}^{T+L},(\blacksquare)^{2T})$ and is immune to possible cancellations of common EBRs in  $\textbf{n}^T=\textbf{n}^{T+L}-\textbf{n}^L$. Thus, we obtain \textit{all} the solutions with a fixed number of auxiliary bands, both normal and composite. This guarantees that no optimal solution will ever be missed since we just have to keep solving Eq. \eqref{enum} for increasing numbers ($\mu^L=0,1,\ldots$) of auxiliary bands until we find the first physical solutions, which by construction are necessarily optimal.

The two methods described so far are mathematically straightforward in that both aim to find solutions to the Diophantine equations. The third, \textit{Real-Space Invariants (RSI) based method}, instead of directly solving Eq. \eqref{matrix_eq}, tries to find a collection of EBRs (specified by $\mathbf{n}^T$), which has the same RSIs as the collection of irreps defined by the vector $\mathbf{v}^T$. The concept of RSI can be introduced as follows. A pseudo-orbital configuration for a given set of bands may be adiabatically deformed into another configuration without breaking the symmetries of the system.
For each WP, RSIs can be defined as invariant quantities under any adiabatic process and computed from the multiplicities of pseudo-orbitals.
Symmetry data determines certain linear combinations of RSIs, where nonzero values indicate that at least one pseudo-orbital must occupy the relevant WPs.
Thus, the RSI based method offers a physical insight into which WPs must be occupied by pseudo-orbitals and efficiently solves the Diophantine problem for simple symmetry data.


For illustrative purposes, we will build the TB model of a realistic but simple photonic crystal depicted in the inset of Fig. \ref{fig:panel224}\textcolor{red}{a} with SG No. 224 ($Pn\Bar{3}m$) \footnote{Another example concerning SG $Pm\Bar{3}m$ (No. 221) is developed in the SM \cite{SuppMat} (see also Refs. \cite{cano2021band,hwang2019fragile,van2018higher,perez2015symmetry,watanabe2018space} therein). This case is simpler than the case shown in the main text, and we make use of it to present a more pedagogical construction of the TB model.}.

Such a crystal is a fully connected geometry composed of dielectric rods arranged along the main diagonals of the cube. It is an inversion symmetric structure with a 3-fold axis along the diagonal of the cube, as can be seen in Fig. \ref{fig:panel224}\textcolor{red}{a} (inset).
By analyzing the symmetry properties of the electric field solutions at each HSP, we obtain the irrep labels in Fig. \ref{fig:panel224}\textcolor{red}{a}.
According to our previous discussion, the first step to finding a basis of pseudo-orbitals is to use the irreps at every HPS to write down the transverse symmetry vector $\mathbf{v}^T$. To do so, we must first decide how many bands should be included in the model. In general, this is a nontrivial decision, as TB models cannot be constructed for arbitrary sets of bands. Specifically, the bands described by a TETB model must fulfill the following conditions: 1) The corresponding irreps must satisfy the compatibility relations, and 2) all the topological symmetry indicators must be trivial.

The first condition is automatically satisfied by any isolated set of bands, i.e., by a collection of bands separated by gaps from all others. This happens, for instance, when the lower frequency bands are separated from the rest by a global gap, which in the case of photonic crystals is known as the \textit{fundamental gap}. The second condition is satisfied for electronic bands or 1D and 2D PhCs bands if the irreps at the HSP can be written as a sum of EBRs. However, for 3D PhCs, the second condition is harder to check owing to the ill-defined irrep content for the zero-frequency modes at $\Gamma$. 

Looking at Fig. \ref{fig:panel224}\textcolor{red}{a}, no global gap that one could use to easily define a suitable set of bands for the TETB model is appreciated. Consequently, we must artificially detach a set of lower frequency bands from the rest to satisfy conditions 1) and 2) in the previous paragraph. This is a highly nontrivial problem which, fortunately, has been completely solved in Ref. \cite{christensen2022location}, where the minimum \textit{transverse connectivities} $\mu_1^T$ for the 230 space groups are computed.
$\mu_1^T$ is the minimum number of bands that an isolated set must contain, including the zero frequency transverse singularity, to satisfy the compatibility relations. Moreover, all possible transverse symmetry vectors, including bands up to the fundamental gap, are tabulated in Ref. \cite{christensen2022location}, together with their topological symmetry indicators. One can also consider \textit{second minimal solutions}, which contain $\mu_2^T$ bands up to the fundamental gap and, in general, n-th minimal solutions with $\mu_n^T$ bands~\footnote{The corresponding symmetry vectors can be computed using the Julia package developed by T. Christensen: PhotonicBandConnectivity \cite{PhotonicBandConnectivity}}. 

In the case of $SG224$ with TRS, Ref. \cite{christensen2022location} gives $\mu_1^T=4$, with $4$ possible transverse symmetry vectors, and $\mu_2^T=6$ with $22$ different transverse symmetry vectors. This means we can build TETB models with four or six transverse bands but not with 1, 2, 3, or 5. 
There are, of course, other possibilities with more bands corresponding to $\mu_n^T$ for $n > 2$ that will not be considered here. The irreps labeled in Fig. \ref{fig:panel224}\textcolor{red}{a} don’t correspond to any of the four minimal solutions. Still, one of the $22$ second minimum solutions precisely matches the irreps for the six lower frequency bands, namely
\begin{align} \label{eq:symVec224}
    \mathbf{v}^T = [&(\blacksquare)^{2T}+\Gamma_2^-+\Gamma_4^-,R_4^-+R_5^+, \nonumber \\
    &M_1+2M_4,X_1+X_3+X_4].
\end{align}
The second condition is also satisfied, as all the symmetry indicators for $SG224$ with TRS are trivial \cite{po2017symmetry}. This means that building a TETB model that faithfully replicates the properties of the six lower frequency bands in the spectrum should be possible.

To obtain $\mathbf{v}^T$ from a linear combination of EBRs, we apply the three methods mentioned above (the details are shown in the SM \cite{SuppMat} (see also Refs. \cite{cano2021band,hwang2019fragile,van2018higher,perez2015symmetry,watanabe2018space} therein)), which in this case lead to the unique minimal solution 
\begin{equation}\label{eq:GEBR}
\textbf{n}^T = A_{2u}@4b + A_{2u}@4c - A_{1}@2a.
\end{equation}
Note that the RSI method yields nonzero RSIs for WPs $2a,4b,4c$ for $\textbf{v}^T$ in Eq.~\eqref{eq:symVec224}, which are precisely the WPs involved in the minimal solution in  Eq.~\eqref{eq:GEBR}. See the SM \cite{SuppMat} (see also Refs. \cite{cano2021band,hwang2019fragile,van2018higher,perez2015symmetry,watanabe2018space} therein) for the detailed computation of RSIs.

According to our previous analysis, the positive multiplicity EBRs in the solution $A_{2u}@4b + A_{2u}@4c$ ($=\textbf{n}^{T+L}$) determine the locations of the pseudo-orbitals that can be used to build a reliable TETB model, while the negative multiplicity  
$A_1@2a$ ($=\textbf{n}^L$) describes the auxiliary bands. Thus, a TETB model can be constructed solely from two EBRs of dimension four, $A_{2u}@4b$, and $A_{2u}@4c$. This means that we will need eight pseudo-orbitals to build a TETB model with six transverse physical bands and two longitudinal auxiliary bands.
Later, we will prove that spectral filtering can remove the auxiliary longitudinal bands $\mathbf{v}^L$ by tuning the TETB parameters \textit{a posteriori}. Note also that using the function \textit{BandRep} at the BCS~\cite{elcoro2017double} to obtain the irrep content of the EBRs at $\Gamma$,  Eq.~\eqref{BS} yields $(\blacksquare)^{2T}=\Gamma_4^- -\Gamma_1^+$, which in this case agrees with the particular surrogate representation tabulated in Ref.~\cite{christensen2022location}. 

\subsection{Mapping to a Photonic TETB} \label{subsec: mapping}

Once a set of candidate pseudo-orbitals has been determined following any of the three methods, we construct a TETB model based on them.
We look for a TETB model that satisfies the following conditions:
the additional degrees of freedom introduced as longitudinal modes ($\mathbf{v}^L$) represent energy bands away from the physical bands,
the TETB model captures the features of the transverse bands ($\mathbf{v}^T$) in the energy window of interest, 
and the model reproduces the $(\blacksquare)^{2T}$ obstruction at $\Gamma$ and all the symmetry, topology, and energetic features of the active bands in the PhC. 

To proceed, we exploit a formal mapping between the cell-periodic Schrödinger and Maxwell wave equations, which are linear and quadratic in time, respectively.
This allows us to relate the energy $\mathcal{E}$ of the electronic wavefunction $\phi(\mathbf{r})$ in presence of a crystal periodic potential $V(\mathbf{r})$

\begin{equation}
    \left[ \frac{-\hbar^2}{2m} \grad^{2} + V(\mathbf{r}) \right] \phi(\mathbf{r}) = \mathcal{E} \phi(\mathbf{r}),
    \label{schrodinger}
\end{equation}
and the frequency of light $\omega$ in media with periodic dielectric permittivity $\varepsilon(\mathbf{r})$

\begin{equation}
    \grad \times \left( \frac{1}{\varepsilon(\mathbf{r})} \grad \times \mathbf{H}(\mathbf{r}) \right) = 
    \left( \frac{\omega}{c} \right)^{2} \mathbf{H}(\mathbf{r}),
    \label{maxwell}
\end{equation}
according to $\mathcal{E}\sim\omega^2$ \cite{de2018}.

This quadratic mapping allows us to construct an effective solid-state inspired optical 3D TETB model by enforcing the eigenvalues ($\mathcal{E}$) of the set of transversal bands to be positive, $\mathcal{E}=\omega^2 \geq 0$, while the lowest set of longitudinal bands to be negative, $\mathcal{E}=\omega^2 \leq 0$.
Note that since the frequency of the electromagnetic solutions is $\omega=\sqrt{\mathcal{E}}$, the final real spectra will not contain the auxiliary nonphysical modes.
Forcing the longitudinal modes and active transverse modes to be isolated from each other except at $\mathcal{E}=0$ enables us to achieve all the previous points.

We proceed via a four-step strategy to construct a reliable TETB that satisfies the abovementioned constraints.

\begin{enumerate}
    \item From the EBR decomposition obtained by any of the three methods described in Section A, we identify  $\textbf{n}^{T+L}$ and $\textbf{n}^L$.
    \item We build a TETB model with generic free parameters from a set of pseudo-orbitals with the symmetries and Wyckoff positions dictated by the EBRs in $\textbf{n}^{T+L}$.
    \item We analyze the symmetry content of the bands induced by these orbitals and identify which modes can be associated uniquely to $\textbf{v}^L$ and $\textbf{v}^T$. 
    \item We tune the parameters of the TETB (onsite energies and hoppings) so that the set of longitudinal bands become the lower negative energy ($\mathcal{E}=\omega^2 \leq 0$) modes.
    Then, we fit the $\mathcal{E}=\omega^2\geq0$ bands to the square of the electromagnetic frequencies obtained numerically for the PhC.
\end{enumerate}
This will result in a Transversality-Enforced Tight-Binding (TETB) model, with Hamiltonian $H(\textbf{k})$, which captures all the symmetry, topology, and energetic features of the transverse bands in the photonic crystal. It is essential to clarify that the eigenstates in this model do not represent the physical, vectorial electromagnetic fields. Instead, they offer an effective description in terms of a scalar wavefunction capable of successfully capturing the boundary-localized electromagnetic energy concentration in finite systems that presents the same momentum-space topological invariants as the real PhC \cite{devescovi2023axion}.
\begin{figure*}
    \centering
    \includegraphics[width=\textwidth]{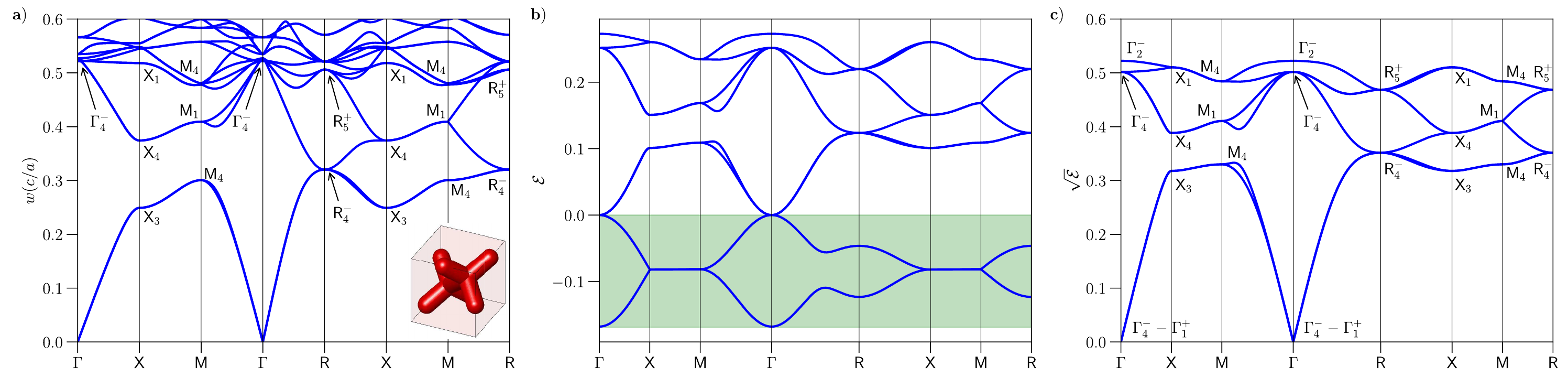}
    \caption{Band structures obtained from MPB and TETB model for a crystal with SG $Pn\Bar{3}m$ (No. 224) without an external magnetic field applied.
    \textbf{a)} Dielectric structure of a crystal with SG $Pn\Bar{3}m$ (No. 224) with its associated band structure \cite{devescovi2021cubic}.
    \textbf{b)} Band structure obtained from the TETB model.
        The bands enclosed by the green-shaded region belong to the additional modes included to regularize the symmetry content at $\Gamma$.
    \textbf{c)} Band structure obtained from the TETB model after applying the mapping to the electromagnetic-wave equation frequency and imposing the transversality constraint.}
    \label{fig:panel224}
\end{figure*}

To illustrate the strategy, we consider again the example in Fig. \ref{fig:panel224}\textcolor{red}{a} with SG No. 224 and symmetry vector in Eq.~\eqref{eq:symVec224}. 
According to Eq.~\eqref{eq:GEBR}, we can build a TB model by placing pseudo-orbitals that transform as $A_{2u}$ in WPs 
 $4c:(1/2,1/2,1/2)$ and $4b:(0,0,0)$, both with site-symmetry group $\bar{3}m$.
This model gives rise to a TETB Hamiltonian, which can be expressed as a $8\times 8$ matrix $H(\textbf{k})$, where the symmetry of the crystal constrains the functional dependence on $\textbf{k}$. Invariance under any space group transformation $g$ imposes the following constraint on $H(\textbf{k})$
\begin{equation}\label{eq:con1}
    gH(\textbf{k})g^{-1}=H(g\textbf{k}),
\end{equation}
where $g\textbf{k}\!\equiv\! R\textbf{k} $ for $g\!=\!\{ R | \textbf{v}\}$.
Similarly, invariance under TRS implies
\begin{equation}\label{eq:con2}
    H(-\mathbf{k})=H^*(\mathbf{k}),
\end{equation}
where $^*$ denotes complex conjugation.
Including interactions up to third-nearest neighbors, the Hamiltonian can be written in terms of nine independent real parameters:
$\alpha_1$ and $\alpha_2$ are on-site energies, while the six remaining parameters are first ($a_1$), second ($a_2$, $r_2$, $s_2$, $w_2$) and third ($a_3$,$r_3$)
nearest neighbor hoppings. The resulting TETB Hamiltonian $H(\textbf{k})$ is given in the SM \cite{SuppMat} (see also Refs. \cite{cano2021band,hwang2019fragile,van2018higher,perez2015symmetry,watanabe2018space} therein).

The free parameters in the Hamiltonian are fitted to the band structure of the crystal, shown in Fig. \ref{fig:panel224}\textcolor{red}{a}, while at the same time forcing the auxiliary nonphysical longitudinal bands to have negative eigenvalues. The resulting parameters are given in Table \ref{tab:para}. 
\begin{table}
    \centering
    \begin{tabular}{cc}
        Parameters & Value \\\hline\hline
        $\alpha_1$ & 0.306159 \\
        $\alpha_2$ & 0.123753 \\
        $a_1$ & 0.157927 \\
        $a_2$ & -0.064197 \\
        $r_2$ & 0.068848 \\
        $s_2$ & -0.100771 \\
        $w_2$ & 0.022712 \\
        $a_3$ & 0.035449 \\
        $r_3$ & -0.041474 \\
    \end{tabular}
    \caption{Parameters of the TETB Hamiltonian for SG $Pn\Bar{3}m$ (No. 224).}
    \label{tab:para}
\end{table}
We perform the fitting using a least-square minimization routine at all HSPs.
The cost function is a multiobjective, multivariable function, which measures the distances between the square of the frequencies computed numerically in MPB for the PhC and the eigenvalues of the TETB for each irrep.
The objective vector contains the multiple distances, while 
the variables are the TETB coefficients.
Specifically, we use the least-squares function of the \textit{scipy.optimize} package in Python \cite{least_squares}.

The eigenvalues of the resulting TETB model are shown in Fig. \ref{fig:panel224}\textcolor{red}{b}. 
The longitudinal bands are in the green-shaded region and represent the negative eigenvalues, detached from the physical, transverse ones.
Finally, we take the square root of the eigenvalues ($\mathcal{E}=\omega^2$) and discard the longitudinal bands, obtaining the physical band-structure shown in Fig. \ref{fig:panel224}\textcolor{red}{c}.
One can see that the TETB model reproduces the MPB band structure of the crystal accurately for the six lowest energy bands, both in their dispersion and symmetry content.
Another example, based on SG $Pm\Bar{3}m$ (No. 221), is presented in the SM \cite{SuppMat} (see also Refs. \cite{cano2021band,hwang2019fragile,van2018higher,perez2015symmetry,watanabe2018space} therein), which includes a detailed step-by-step derivation of the procedure.

\subsection{Introducing a Magnetic Field into the Model}

Some of the essential applications of topological photonics rely on Time Reversal Symmetry (TRS) breaking since it stabilizes strong topology in the Cartan–Altland–Zirnbauer (CAZ) ten-fold classification of topological materials \cite{altland1997nonstandard,schnyder2008classification}.
Usually, TRS breaking in PhCs is achieved using gyroelectric or gyromagnetic materials and applying an external static magnetic field or equivalently through intrinsic remnant magnetization.
To mimic such effects, we develop a general method to simulate the interaction of an external and static magnetic field with a PhC in our TETB models.

One possible approach is to extend the symmetry analysis previously conducted without a magnetic field by incorporating magnetic SGs. However, this method is inherently dependent on the direction of the magnetic field and must be repeated each time the field's orientation changes. To overcome this limitation, we sought to develop a more versatile approach—one that eliminates the need to construct a new TB model for each case, allowing us to reuse the previously established model while accommodating an arbitrarily oriented magnetic field.

In conventional TB models for electronic band structures, the influence of a static magnetic field is typically introduced via the minimal coupling of electrons to the vector potential $\mathbf{A}$, often implemented through the Peierls substitution \cite{peierls1933theorie, luttinger1951effect}. However, this mechanism is inapplicable to photonic systems due to the charge neutrality of photons.

This forces us to use non-minimal couplings, 
where the magnetic field $\textbf{H}$ is treated as a perturbation in the system's response.
Accordingly, the Hamiltonian, including the effects of a magnetic field, can be represented by

\begin{equation}
    H_M (\textbf{k},\textbf{H}) = H(\textbf{k}) + f(\textbf{k},\textbf{H}),
\end{equation} 
where $H(\textbf{k})$ is the TETB Hamiltonian of the previous section 
and $f(\mathbf{k},\textbf{H})$ is a magnetic field depending perturbation.
In most cases, taking $f(\textbf{k},\textbf{H})$ as a linear function in the components of $\textbf{H}$ is enough to model the effects of the magnetic field. That's what we do in this paper, 
but higher orders of $\textbf{H}$ could easily be added to model the effects of higher magnetic field intensities if necessary.

The constraints of Eqs. \eqref{eq:con1} and \eqref{eq:con2} must be generalized to include the magnetic field and become 
\begin{equation}\label{eq:conB1}
    gf(\textbf{k},\textbf{H})g^{-1}=f(g\textbf{k},g\textbf{H})
\end{equation}
for any space group transformation $g$ and 
\begin{equation}\label{eq:conB2}
    f(-\mathbf{k},-\mathbf{H})=f^*(\mathbf{k},\mathbf{H})
\end{equation}
for TRS. Here, we would like to comment on the following subtlety, which sometimes is a source of confusion. Even though coupling to an external magnetic field breaks the TRS of the crystal, we must still impose invariance under joint transformations of the crystal \textit{plus} the magnetic field. This is true not only for TRS but also for the original space group transformations. 

Coming back to the example of the crystal with SG $Pn\Bar{3}m$ (No. 224), the introduction of a $z$-directed gyromagnetic bias gives rise to a topological charge-1 Weyl dipole oriented along the $k_z$ axis \cite{devescovi2021cubic}.
This is due to the triple degeneracy at the $R$-point of the TRS case, making the crystal a Weyl-photonic semimetal.
To model this phenomenon via the TETB, we will use a first-order perturbation in $\textbf{H}$, with the magnetic field directed along the $z$ axis.
As a result of the constraints imposed by Eqs. \eqref{eq:conB1} and \eqref{eq:conB2}, the linear coupling to the magnetic field depends on five free, real parameters:
$\delta_1$ for first-nearest neighbor hopping terms, and 
$\delta_2$, $\beta_2$, $\kappa_2$ and $\epsilon_2$ for second-nearest neighbor hoppings.
Due to symmetry constraints, first-order linear perturbation does not affect the third nearest neighbor hoppings. The corresponding magnetic TETB Hamiltonian $H_M(\textbf{k},\textbf{H})$ is given in the SM \cite{SuppMat} (see also Refs. \cite{cano2021band,hwang2019fragile,van2018higher,perez2015symmetry,watanabe2018space} therein).

\begin{figure*}
    \centering
    \includegraphics[width=\textwidth]{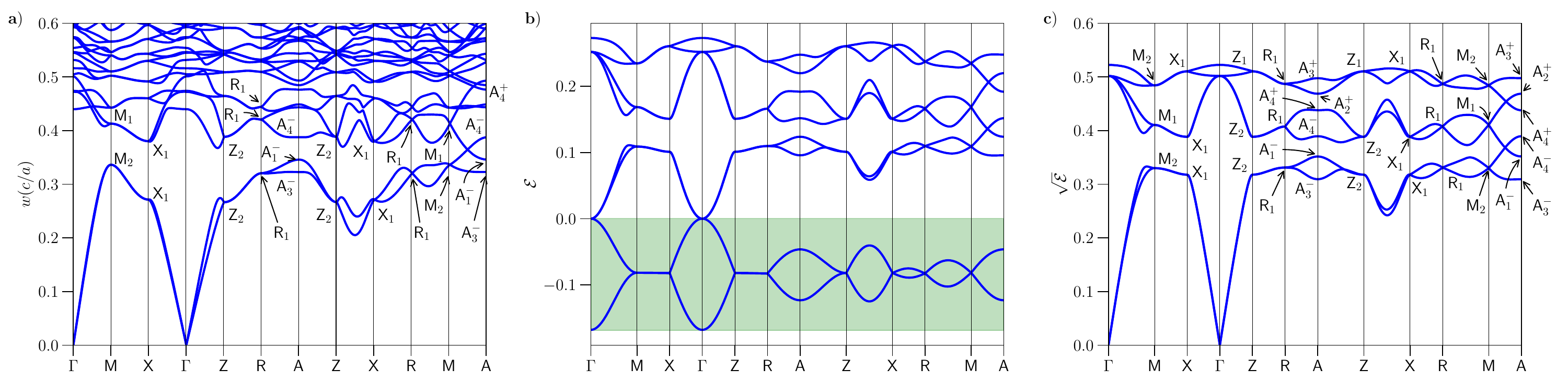}
    \caption{Band structures obtained from MPB and TETB model for a crystal with SG $Pn\Bar{3}m$ (No. 224) with an external magnetic field applied along $z$.
    \textbf{a)} Frequency bands obtained from the crystal with a magnetic field applied along $z$.
    \textbf{b)} Band structure obtained from the TB model with a magnetic field applied along $z$.
        The bands enclosed by the green-shaded region belong to the additional modes included to regularize the symmetry content at $\Gamma$.
    \textbf{c)} Frequency bands obtained from the TB model with a magnetic field applied along $z$.}
    \label{fig:panel224Mag}
\end{figure*}

Now, the new parameters need to be adjusted to fit the MPB band structure of the crystal affected by the static magnetic field (shown Fig. \ref{fig:panel224Mag}\textcolor{red}{a}).
As in the previous section, we apply the same minimization routine over the new parameters, keeping the last TETB parameters (Table \ref{tab:para}) unchanged.
We obtain the values shown in Table \ref{tab:magPara} for the perturbation. This approach - once the magnetic field direction is fixed -  must be consistent with the symmetry analysis using magnetic SGs mentioned at the beginning. A consistency check is performed in the SM \cite{SuppMat} (see also Refs. \cite{cano2021band,hwang2019fragile,van2018higher,perez2015symmetry,watanabe2018space} therein).

\begin{figure}
    \centering
    \includegraphics[width=0.45\textwidth]{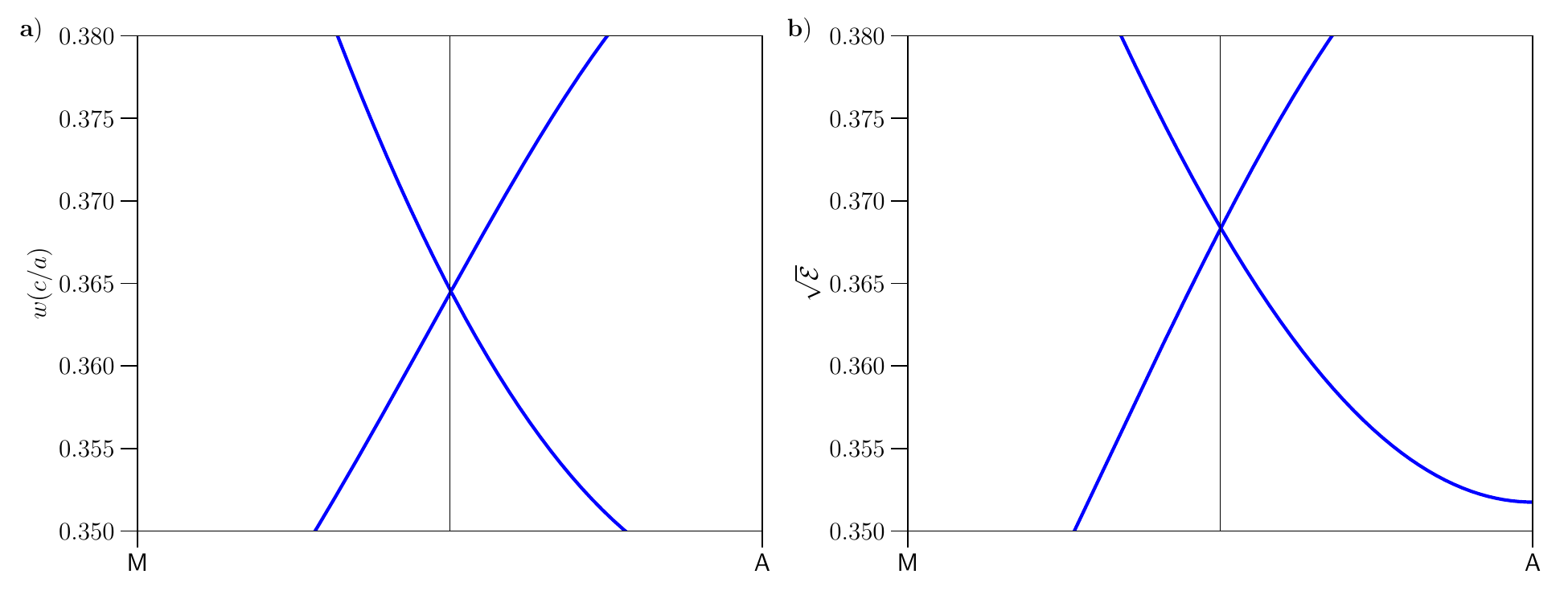}
    \caption{Position of the Weyl point obtained from MPB and TB model for a crystal with SG $Pn\Bar{3}m$ (No. 224) in perfect agreement with Ref. \cite{devescovi2021cubic}.
    \textbf{a)} Weyl point at exactly half the MA line for the MPB computations.
    \textbf{b)} Weyl point situated at exactly half the MA line for the TB model.}
    \label{fig:panelWeyls}
\end{figure}

Fig. \ref{fig:panel224Mag}\textcolor{red}{b} displays the eigenvalues of the perturbed TETB model with a magnetic field along $z$. Again, the longitudinal auxiliary bands are kept in the green-shaded region, showing that they continue to be negative even in the presence of the magnetic field.
\begin{table}
    \centering
    \begin{tabular}{cc}
        Parameters & Value \\\hline\hline
        $\delta_1$ & 0 \\
        $\delta_2$ & $0.001$ \\
        $\beta_2$ & $-0.001$ \\
        $\kappa_2$ & $0.001$ \\
        $\epsilon_2$ & $-0.001$ \\
    \end{tabular}
    \caption{Parameters of the linear function $f_L(\textbf{k},\textbf{H})$.}
    \label{tab:magPara}
\end{table}
Finally, using $\mathcal{E}=\omega^2$, the longitudinal modes are discarded, obtaining the physical band-structure shown in Fig. \ref{fig:panel224Mag}\textcolor{red}{c} that closely matches the one obtained through exact numerical simulations of the PhC.
Note that in the MPB simulations, the applied magnetic field generates a pair of Weyl points \cite{devescovi2021cubic}, one of them halfway along the MA high symmetry line. As shown in Fig \ref{fig:panelWeyls}, 
the TETB model exactly replicates this behavior.

\section{Conclusions} \label{sec: end}

In summary, we propose the first systematic method to construct a reliable TB representation of 3D PhCs in this article, even if maximally localized WFs do not exist for such systems \cite{wolff2013generation}.
We show how this can be achieved by developing a TETB model capable of capturing and regularizing the $\Gamma$-point electromagnetic obstruction that arises due to the transversality constraint of Maxwell's equations while accurately reproducing the symmetries of the PhC.
This method proceeds by adding auxiliary longitudinal modes to the physical transverse bands of the 3D PhC \cite{christensen2022location}. 
We propose three strategies to identify optimal pseudo-orbital candidates for the TETB: two methods that directly solve the original Diophantine system of equations and one method that solves a related but usually simpler system.
The direct approach proceeds via linear optimization routines (optimization method) or constraint solvers (enumeration method), while the third method is based on calculations of RSIs.
Additionally, we developed a code implementing the enumeration method, making it accessible for non-experts on the topic.
We can ultimately discard the nonphysical longitudinal modes by establishing a formal mapping between Maxwell's and Schrödinger's equations.
The resulting TETB accurately reproduces all the symmetry, topology, and energetic dispersion of the transverse bands in the PhC.
Finally, we show how to model gyrotropy by providing a magnetic version of our TETB model using non-minimal coupling.

Our work provides the first systematic method to analytically model the photonic bands of the lowest transverse modes over the entire BZ via a TETB model. 
The lower computational cost of TETB models compared to exact solvers will enable more complex theoretical developments in topological photonics by facilitating the calculation of larger and more intricate supercells.
For example, recently, this method allowed us to simulate the higher-order response on the hinges of photonic axion insulators \cite{devescovi2023axion}.
This development was only possible using TETB models because calculating $xy$-confined rod geometries of 3D supercells was beyond state-of-the-art high-performance computing clusters. Therefore, our TETB will facilitate the study of boundary responses of future photonic topological phases, particularly in the case of 3D PhC.

\textbf{Acknowledgements} \par 

C.D. and A.M.P. have contributed equally to this work.

A.G.E., A.M.P, M.G.D., M.G.V. and C.D. acknowledge support from the Spanish Ministerio de Ciencia e Innovación (PID2022-142008NB-I00).
A.G.E., A.M.P, and C.D. also acknowledge support from the Basque Government Elkartek program (KK- 533 2023/00016) and the Gipuzkoa Provincial Council within the QUAN-000021-01 project.
A.G.E. and M.G.V. acknowledge funding from the IKUR Strategy under the collaboration agreement between Ikerbasque Foundation and DIPC on behalf of the Department of Education of the Basque Government, Programa de Ayuda de Apoyo a los agentes de la Red Vasca de Ciencia, Tecnolog\'ia e Innovaci\'on acreditados en la categor\'ia de Centros de Investigaci\'on B\'asica y de Excelencia (Programa BERC) from the Departamento de Universidades e Investigaci\'on del Gobierno Vasco and Centros Severo Ochoa AEI/CEX2018-000867-S from the Spanish Ministerio de Ciencia e Innovaci\'on.
M.G.V. thanks support to the Deutsche Forschungsgemeinschaft (DFG, German Research Foundation) GA 3314/1-1 – FOR 5249 (QUAST) and  the Canada Excellence Research Chairs Program for Topological Quantum Matter. The work of JLM has been partly supported by the Basque Government Grant No. IT1628-22 and the PID2021-123703NB-C21 grant funded by MCIN/AEI/10.13039/501100011033/ and ERDF; ``A way of making Europe”. The work of B.B. and Y.~H. is supported by the Air Force Office of Scientific Research under award number FA9550-21-1-0131. Y.~H. received additional support from the US Office of Naval Research (ONR) Multidisciplinary University Research
Initiative (MURI) grant N00014-20-1-2325 on Robust
Photonic Materials with High-Order Topological Protection. C.D. acknowledges financial support from the MICIU through the FPI Ph.D. Fellowship CEX2018-000867-S-19-1. M.G.D. acknowledges financial support from the Government of the Basque Country through the pre-doctoral fellowship PRE\_2022\_2\_0044.

\bibliography{references}

\end{document}


\title{Supplementary Material for "Transversality-Enforced Tight-Binding Models for 3D Photonic Crystals aided by Topological Quantum Chemistry"}

\author{Antonio Morales-P\'erez*}
\email{antonio.morales@dipc.org}
\affiliation{Donostia International Physics Center, Paseo Manuel de Lardizabal 4, 20018 Donostia-San Sebastian, Spain.}
\affiliation{Material and Applied Physics Department, University of the Basque Country (UPV/EHU), Donostia-San Sebastián, Spain.}

\author{Chiara Devescovi*}
\affiliation{Donostia International Physics Center, Paseo Manuel de Lardizabal 4, 20018 Donostia-San Sebastian, Spain.}

\author{Yoonseok Hwang}
\affiliation{Department of Physics, University of Illinois at Urbana-Champaign, Urbana, IL, USA}

\author{Mikel García-Díez}
\affiliation{Donostia International Physics Center, Paseo Manuel de Lardizabal 4, 20018 Donostia-San Sebastian, Spain.}
\affiliation{Physics Department, University of the Basque Country (UPV/EHU), Bilbao, Spain}

\author{Barry Bradlyn}
\affiliation{Department of Physics, University of Illinois at Urbana-Champaign, Urbana, IL, USA}

\author{Juan L. Ma\~nes}
\affiliation{Physics Department, University of the Basque Country (UPV/EHU), Bilbao, Spain}

\author{Maia G. Vergniory}
\email{maia.vergniory@cpfs.mpg.de}
\affiliation{Donostia International Physics Center, Paseo Manuel de Lardizabal 4, 20018 Donostia-San Sebastian, Spain.}
\affiliation{Département de Physique et Institut Quantique, Université de Sherbrooke, Sherbrooke, QC J1K 2R1 Canada.}

\author{Aitzol Garc\'ia-Etxarri}
\email{aitzolgarcia@dipc.org}
\affiliation{Donostia International Physics Center, Paseo Manuel de Lardizabal 4, 20018 Donostia-San Sebastian, Spain.}
\affiliation{IKERBASQUE, Basque Foundation for Science, Mar\'ia D\'iaz de Haro 3, 48013 Bilbao, Spain.}

\date{\today}
\maketitle

\section{Solving the Diophantine equations}

Here, we give a detailed account of the three methods outlined in the main text to decompose the symmetry vector $\mathbf{v}^T$ of the physical transverse bands in terms of Elementary Band Representations (EBRs). This can be formulated as a linear problem

\begin{equation}
\textbf{v}^T=\textbf{A}\textbf{n}^T,
\label{matrix_eq}
\end{equation}
where $\textbf{n}^T$ is an integer vector with the multiplicities  of the $N_{EBR}$ EBRs.
As explained in the main text, $\text{rank}(\textbf{A})\leq\min(N_{EBR}, N_{irr})$ in general, and we can, in principle, find infinitely many solutions to Eq. \eqref{matrix_eq}.
A possible way to solve this equation is to compute the Smith decomposition of the integer matrix $\textbf{A}$

\begin{equation}
\label{smithA}
\textbf{A}=\textbf{U}^{-1}\textbf{D}\textbf{V}^{-1},
\end{equation}
where $\textbf{U}$ and $\textbf{V}$ are invertible matrices over the integers 
and $\textbf{D}$ is an integer diagonal matrix whose non-zero entries are called divisors of $\textbf{A}$ \cite{cano2021band}.
Note that $\textbf{D}$ generally contains zero diagonal entries and is thus non-invertible in the usual sense.
Now, the general solution $\textbf{n}^T$ can be computed as

\begin{equation}
\textbf{n}^T = \textbf{V} \textbf{D}^{+} \textbf{U} \textbf{v}^T + (\mathds{1}_{N_{EBR}} - \textbf{V} \textbf{D}^{+} \textbf{U} \textbf{A}) \textbf{z}
\label{pseudoinverse}
\end{equation}
where $\textbf{D}^+$ denotes the pseudo-inverse of $\mathbf{D}$ 
and $\textbf{z}$ is an arbitrary integer vector. 
Eq. \eqref{pseudoinverse} can be derived as follows.
Eqs. \eqref{matrix_eq} and \eqref{smithA} imply $[\textbf{V}^{-1} \textbf{n}^T]_i = [\textbf{D}^+ \textbf{U} \textbf{v}^T]_i$ for $i=1,\dots,r=\text{rank}(\textbf{A})$.
However, $[\textbf{V}^{-1} \textbf{n}^T]_{i'}$ is undetermined for $i'=r+1,\dots, N_{irr}$.
Thus, we define $y_{i'}=[\textbf{V}^{-1} \textbf{n}^T]_{i'}$.
Note that, because $\textbf{V}$ is a unimodular matrix and $\textbf{n} \in \mathbb{Z}^{N_{EBR}}$, $y_{i'} \in \mathbb{Z}$.
Thus, we equate $\textbf{V}^{-1} \textbf{n}^T = \textbf{D}^+ \textbf{U} \textbf{v}^T + (\mathds{1}_{N_{EBR}}-\textbf{D}^+ \textbf{D}) \textbf{y}$.
Here, we define $\textbf{y}$ such that $(\textbf{y})_{i'}=y_{i'}$ and integer-valued free parameters otherwise.
By multiplying $\textbf{V}$ and introducing $\textbf{z}=\textbf{V}\textbf{y}$, Eq. \eqref{pseudoinverse} is obtained.
Eq.~\eqref{pseudoinverse} represents the general solution to Eq. \eqref{matrix_eq} parametrized by the $N_{irr}$-dimensional integer vector $\textbf{z}$.

As discussed in the main text, different solutions will correspond to TETB models with different numbers $\mu^L$ of auxiliary bands, but for practical reasons, we are mainly interested in ``optimal'' solutions where that number is minimal. 
The easiest way to solve Eq. \eqref{matrix_eq} is to set $\mathbf{z}=0$ in Eq. \eqref{pseudoinverse}.
This provides a solution but is not guaranteed to be optimal. We can illustrate this fact by applying  Eq.~\eqref{pseudoinverse} to the symmetry vector for the PhC with SG224 considered in the main text
\begin{equation}\label{vT} 
    \mathbf{v}^T = [(\blacksquare)^{2T}+\Gamma_2^-+\Gamma_4^-,R_4^-+R_5^+,M_1+2M_4,X_1+X_3+X_4],
\end{equation}
where we take $(\blacksquare)^{2T} = \Gamma_4^- - \Gamma_1^+$. This yields 
\begin{equation}
    \textbf{n}^T = A_1@2a + 2T_{2}@2a - A_{1g}@4b - A_{1g}@4c,
\end{equation}
where the positive component of the solution,  $\textbf{n}^{T+L}=A_1@2a + 2T_{2}@2a$, defines a TETB model obtained by placing orbitals at WP $2a$. Taking into account that $A_1$ and $T_2$ are one and three-dimensional irreps, respectively, the TETB model describes a total of $1\times 2+2\times 3\times 2 = 14 $ bands (orbitals), that include the six transverse bands in \eqref{vT} plus 
eight $(=1\times 4+1\times 4)$ auxiliary bands associated to the negative component $\textbf{n}^{L}=A_{1g}@4b + A_{1g}@4c$. This is a far cry from the minimal TETB model with just two auxiliary bands obtained by application of the methods introduced in the main text, which will be discussed later in this section.

In this paper, we propose three strategies to find solutions to Eq. \eqref{matrix_eq} that minimize the number of bands in the TETB model. In Sec. \ref{subsec:numerical}, we give two approaches that directly solve the Diophantine equation \eqref{matrix_eq}. The first uses optimization techniques, taking the number of bands as an objective function to minimize, while the enumeration method obtains all possible solutions with a given, fixed number of bands. These two methods are mathematically straightforward to follow but require running computational constraint solvers or minimization algorithms due to the usually large number of linear equations in \eqref{matrix_eq}. In Sec. \ref{subsec:RSI}, we take a different approach where we use Real-Space Invariants (RSI) to define an auxiliary problem that, in certain situations, can be easier to solve than Eq. \eqref{matrix_eq}. In some cases, the third method directly outputs the minimal solution but requires knowledge of the symmetry constraints of irreps at $\Gamma$ \cite{watanabe2018space,christensen2022location} and uses state-of-the-art group theory arguments. 

Unlike electronic systems, where every solution to Eq. \eqref{matrix_eq} can be used in principle to build a valid TB model, in the case of photonic crystals, we must impose additional constraints on the solutions. As these are related to the existence of auxiliary bands and the $\Gamma$ point singularity,  we must take a brief detour and discuss the role played by auxiliary bands before we embark on the study of the three methods.

\begin{figure*}
    \centering
    \vspace*{-2.5cm}
    \includegraphics[width=\textwidth]{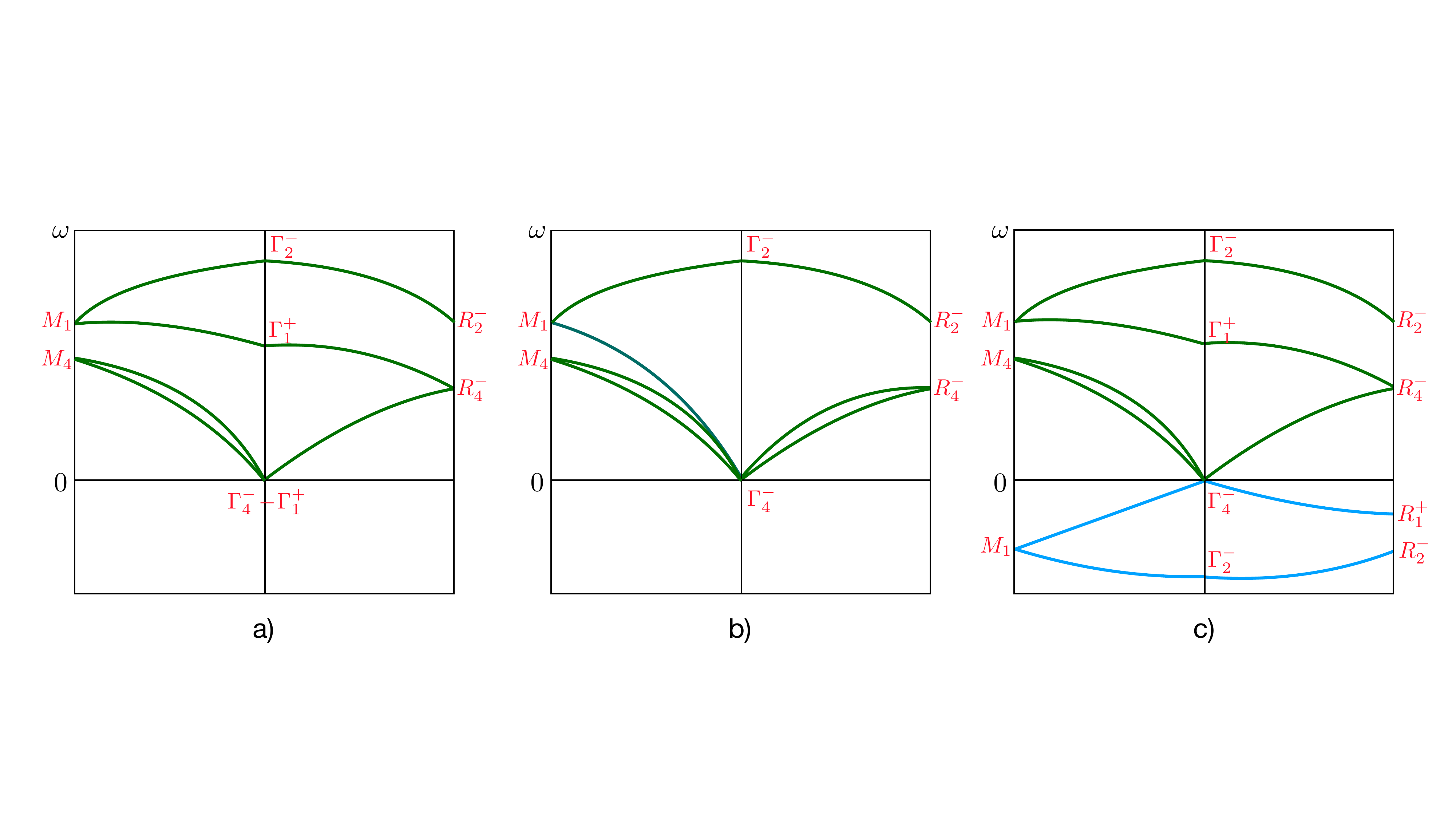}
    \vspace*{-2cm}
    \caption{Band structures for 
    \textbf{a)} the PhC with symmetry vector given by Eq. \eqref{PhC},
    \textbf{b)} the TB model without auxiliary bands given by Eq. \eqref{TB}, and
    \textbf{c)} the optimal TETB with two auxiliary bands defined by Eq. \eqref{opt}. This representation is purely schematic and the auxiliary bands in blue are understood to have imaginary frequencies. }
    \label{fig:unphysical}
\end{figure*}

\subsection{Unphysical solutions and the role of auxiliary bands in TETB models}\label{subsec:role}

In order to understand one of the roles played by the auxiliary bands we consider a PhC with the following symmetry vector for SG224~\footnote{All symmetry vectors with minimal transverse connectivity $\mu_1^T=2$ used in this SM have been taken from Ref. \cite{christensen2022location}, while those with second minimal connectivity $\mu_2^T=6$ have been computed using the Julia package developed by T. Christensen: PhotonicBandConnectivity \cite{PhotonicBandConnectivity}} 
\begin{equation}\label{PhC}
    \textbf{v}^T = [(\blacksquare)^{2T} + \Gamma_1^+ + \Gamma_2^-, R_2^- + R_4^-, M_1 + M_4, X_1 + X_3].
\end{equation}
Assuming $(\blacksquare)^{2T} = \Gamma_4^- - \Gamma_1^+$, it is easy to check  that $\textbf{n}^T = A_{2u}@4b$ solves Eq. \eqref{matrix_eq}. As this solution has no negative EBR multiplicities, we could be tempted to think that it is possible to build a TETB model without auxiliary bands.  However, using the \textit{BandRep} application at the BCS, we see that the irrep content of $\textbf{n}^T = A_{2u}@4b$ is given by 
\begin{equation}\label{TB}
\textbf{A}\textbf{n}^T=[\Gamma_4^-+\Gamma_2^-, R_2^- + R_4^-, M_1 + M_4, X_1 + X_3].
\end{equation}
Note that, according Eq. \eqref{PhC}, there is a non-zero frequency state at $\Gamma$ transforming as
$\Gamma_1^+$, which by Eq. \eqref{TB} would be absent in the TB model. In the TB model, the negative multiplicity irrep in $(\blacksquare)^{2T} = \Gamma_4^- - \Gamma_1^+$ has canceled against the non-zero frequency $\Gamma_1^+$ irrep in the transverse bands, with the result that there are now \textit{three} bands with vanishing frequencies, $\omega\to 0$, as $k\to 0$. Of course, this cancellation can never occur in the PhC bands \eqref{PhC}, for the two $\Gamma_1^+$ states have different frequencies. The effect on the TB bands is schematically illustrated in Fig.~\ref{fig:unphysical}\textcolor{red}, which compares the PhC bands~(a) with those in the TB model (b). Thus, this TB model is not transversality-enforced and should be discarded.  

On the other hand, applying the enumeration method (to be discussed in the following subsection) to the symmetry vector in Eq. \eqref{PhC} gives the following optimal solution

\begin{equation}\label{opt}
    \textbf{n}^{T+L} = A_{2u}@4b+A_1@2a\;,\;\;\; \textbf{n}^{L} = A_1@2a \;,\;\;\; (\blacksquare)^{2T} = \Gamma_4^- - \Gamma_1^+
\end{equation}
which includes two auxiliary bands described by the EBR $A_1@2a=\textbf{n}^{L}$ with irrep content given by
\begin{equation}\label{aux}
\textbf{A}\textbf{n}^L=[\Gamma_1^++\Gamma_2^-, R_1^++R_2^- , M_1 , X_1].
\end{equation}

Now, as shown in Fig.~\ref{fig:unphysical}\textcolor{red}{c}, the negative multiplicity irrep in $(\blacksquare)^{2T}$ cancels against the $\Gamma_1^+$ in the auxiliary bands and, unlike the previous case, the finite transverse frequency $\Gamma_1^+$ is preserved.

Note that this is an example of a ``composite solution'', defined in the main text, where the optimal TETB model for a particular crystal has at least one  EBR which is common to $\textbf{n}^{T+L}$ and $\textbf{n}^L$. Therefore, at least in this case, the function of the auxiliary bands is to cancel negative multiplicities in the surrogate representation $(\blacksquare)^{2T} = \Gamma_4^- - \Gamma_1^+$. Thus, in order to get a physically sound TETB model, the following constraint must be satisfied by the corresponding solution to Eq. \eqref{matrix_eq}: \textit{Any negative multiplicity irrep in the surrogate representation must also be present in the auxiliary bands to be mutually canceled}. This guarantees that all the physical transverse irreps in the PhC will also be present in the TETB model.

In this example, we have taken the surrogate representation  $(\blacksquare)^{2T} = \Gamma_4^- - \Gamma_1^+$, but SG224 is one of the 117 space groups mentioned in the main text where the zero frequency irrep content at $\Gamma$ is \textit{unpinned}, and there are infinitely many solutions to the compatibility relations. This suggests the possibility that a better choice without negative multiplicities would render the auxiliary bands unnecessary. To check this possibility, we may use Tables \textcolor{red}{S7} and \textcolor{red}{S8} in the SM of Ref. \cite{christensen2022location}, where the general solution to the compatibility relations for all the unpinned groups is explicitly given. Specifically, for SG224 (and any other space group with point group $m\bar 3 m$)

\begin{equation}\label{surro}
\begin{aligned}
(\blacksquare)^{2T} =-\Gamma_1^++\Gamma_4^- &+l\,(\Gamma_2^+-\Gamma_2^-+\Gamma_5^+-\Gamma_5^-) \\
&+p\,(\Gamma_1^+-\Gamma_1^--\Gamma_2^++\Gamma_2^-+\Gamma_4^+-\Gamma_4^--\Gamma_5^++\Gamma_5^-) \\
&+q\,(-\Gamma_1^++\Gamma_1^--\Gamma_2^++\Gamma_2^-+\Gamma_3^+-\Gamma_3^-),
\end{aligned}
\end{equation}
where $\{l,p,q\}$ are arbitrary integers. It is easy to check that no choice of the three integers exists for which all the multiplicities are positive. Actually, we have checked that this is the case for all 117 unpinned space groups. Thus, at least for the unpinned space groups, the use of auxiliary bands is unavoidable.

The situation differs for some of the 113 pinned groups since, according to Ref. \cite{christensen2022location},  the surrogate representation is strictly positive for the 68 space groups in Table \ref{tab:pinnedSGs}. However, this doesn't necessarily mean that we can construct viable TETB models without auxiliary bands for these groups. The reason is that auxiliary bands accomplish additional goals beyond the cancellation of negative irreps in the surrogate representation. The following example may help clarify the issue. For SG99, for which the surrogate representation is uniquely given by $(\blacksquare)^{2T}=\Gamma_5$, consider the symmetry vector 
\begin{equation}\label{sg99ex1}
    \textbf{v}^T = [(\blacksquare)^{2T}  ,A_5,M_5,Z_5,R_3+R_4,X_3+X_4].
\end{equation}
A look at the application \textit{BandRep} at  the BCS shows that Eq. \eqref{matrix_eq} is solved by 
$\textbf{n}^T = E@1a$, and the TETB model built by placing two $E$ orbitals at the $1a$ WP will reproduce the correct irrep content for the two transverse bands in the PhC without the need of auxiliary bands. 

On the other hand, if for the same group, we consider instead the symmetry vector
\begin{equation}\label{sg99ex2}
    \textbf{v}^T = [(\blacksquare)^{2T}  ,A_2+A_3,M_2+M_3,Z_5,R_3+R_4,X_3+X_4],
\end{equation}
we find no solutions without auxiliary bands. There are, however, two optimal solutions with one auxiliary band each
\begin{equation}\label{optex2}
\begin{aligned}
    \textbf{n}^{T+L} &= A_2@1b + B_2@2c\;,\;\;\; \textbf{n}^{L} = A_2@1a \;,\;\;\; (\blacksquare)^{2T} = \Gamma_5,\\
     \textbf{n}^{T+L} &= A_1@1b + B_1@2c\;,\;\;\; \textbf{n}^{L} = A_1@1a \;,\;\;\; (\blacksquare)^{2T} = \Gamma_5.
    \end{aligned}
\end{equation}
We see that in this case, the auxiliary bands are not required to cancel negative irreps. Instead, their existence shows that the symmetry vector in Eq. \eqref{sg99ex2} can not be written as a sum of EBRs, very much like electronic bands with fragile topology. In other words, for any space group in Table \ref{tab:pinnedSGs}, we may expect to find symmetry vectors to which a TETB model without auxiliary bands can be associated, while this will be impossible for other vectors.

However, even for symmetry vectors that allow the construction of a TETB model without auxiliary bands, these bands are still necessary if we want to construct a model in which it is in principle possible to capture the $\mathbf{k}\cdot\mathbf{p}$ theory of Maxwell's equations at long wavelengths. In particular, we know from Maxwell's equations that in a photonic system, the modes near $\mathbf{k},\omega \rightarrow 0$ are transverse polarized to leading order in $|\mathbf{k}|$ \footnote{This will be strictly true for $k$ along high symmetry directions. See the $\mathbf{k}\cdot\mathbf{p}$ model at the end of section~\ref{sec:NM_221}}. This can only be captured in models with appropriate auxiliary bands. For that reason, the possibility of adding the appropriate auxiliary bands to make a model capable of capturing the $\mathbf{k}\cdot\mathbf{p}$ theory of Maxwell's equations has been included as an option in the code that implements the enumeration method.

The condition to be satisfied by the additional auxiliary bands in any of the space groups in Table \ref{tab:pinnedSGs} is easy to state: \textit{The symmetry vector for the auxiliary bands must include either the trivial (scalar), or the pseudoscalar representation}. Objects transforming according to the pseudoscalar representation are invariant under pure rotations and change signs under ``improper'' operations (inversion, mirrors, and roto-inversions). This explains the structure of the particular surrogate representation in Ref. \cite{christensen2022location}, which is obtained by subtracting the  
scalar from the vector representation ($V$) in an electric field description or 
the pseudoscalar representation from the axial vector representation ($A$) in the magnetic case.
The reason is that the surrogate representation has to be compatible with the transformation properties of the transverse fields for $\mathbf{k},\omega \to 0$, and these are obtained by subtracting the longitudinal component, which for electric fields transforms like a scalar under the little group of $\mathbf{k}$, or a pseudoscalar for magnetic fields. 

For SG99, and for any space group with point group $4mm$, the trivial and pseudoscalar representations are given by $\Gamma_1$ and $\Gamma_4$, respectively. Using $V=\Gamma_1+\Gamma_5$ and $A=\Gamma_4+\Gamma_5$ yields
\begin{equation}\label{sur99}
   (\blacksquare)^{2T}=V-\Gamma_1 = (\Gamma_1+\Gamma_5)-\Gamma_1=A-\Gamma_4 = (\Gamma_4+\Gamma_5)-\Gamma_4=\Gamma_5
\end{equation} 
in agreement with Table \ref{tab:pinnedSGs}. Now, any $\mathbf{k}\cdot\mathbf{p}$ model will describe the complete electric or magnetic fields, not just their transverse components, and for that purpose, the auxiliary bands must supply the subtracted longitudinal components transforming like the scalar or pseudoscalar representations. Thus, the symmetry vector for the auxiliary bands must include either the trivial (scalar) or the pseudoscalar representation, as stated above. Note also that if we write the surrogate representation as $V-\Gamma_1$ or $A-\Gamma_4$ without canceling irreps, this condition is just a particular case of the one previously formulated: \textit{Any negative multiplicity irrep in the surrogate representation must also be present in the auxiliary bands to be mutually canceled.} Some examples are given in the next subsection.

\begin{table}[h!]
    \centering
    \begin{tabular}{lr|c} \hline\hline
        PG & $(\blacksquare)^{2T}$ & SGs \\\hline
        1 & $2\Gamma_1$ & \{1\} \\
        2 & $2\Gamma_2$ & \{3, 4, 5\} \\
        m & $\Gamma_1 + \Gamma_2$ & \{6, 7, 8, 9\} \\
        mm2 & $\Gamma_3 + \Gamma_4$ & \{25,\dots, 46\} \\
        4 & $\Gamma_3\Gamma_4$ & \{75,\dots, 80\} \\
        4mm & $\Gamma_5$ & \{99,\dots, 110\} \\
        3 & $\Gamma_2\Gamma_3$ & \{143,\dots, 146\} \\
        3m & $\Gamma_3$ & \{156,\dots, 161\} \\
        6 & $\Gamma_4\Gamma_6$ & \{168,\dots, 173\} \\
        6mm & $\Gamma_6$ & \{183,\dots, 186\} \\\hline\hline
    \end{tabular}
    \caption{Pinned space groups with positive surrogate representation.}
    \label{tab:pinnedSGs}
\end{table}

\subsection{Direct solution methods} \label{subsec:numerical}

By direct methods, we mean those that attempt to directly solve the original Diophantine system of equations~\eqref{matrix_eq} or, rather, its $\Gamma$-agnostic equivalent 

\begin{equation}
\textbf{v}^T_{(BZ-\Gamma)} = \textbf{A}_{(BZ-\Gamma)} \textbf{n}^T,
\label{diophantine}
\end{equation}
where $\textbf{A}_{(BZ-\Gamma)}$ has been obtained from $\textbf{A}$ by eliminating the irreps at the $\Gamma$ point from the EBR matrix $\textbf{A}$. The chief advantage of using \eqref{diophantine} is that, as explained in the main text, solving it is equivalent to solving Eq. \eqref{matrix_eq} for all possible choices of the surrogate representation. This is important since, as shown in the examples below, the minimum number of auxiliary bands in a solution depends critically on the form of the surrogate representation $(\blacksquare)^{2T}$ which, for the 117 unpinned spaced groups, is only partially constrained by the compatibility relations. By avoiding a commitment to a particular form of the surrogate representation, we can always obtain the solution with the minimum number of auxiliary bands and the most economical TETB model. Actually, rather than an input, the surrogate representation is part of the output of $\Gamma$-agnostic methods and can be obtained as

\begin{equation}
(\blacksquare)^{2T}= \textbf{v}^{T+L}-\textbf{v}^L-\tilde{\textbf{v}}^T,
\label{BS} 
\end{equation}
where $\textbf{v}^L=\textbf{A}\textbf{n}^L$ gives the symmetry vector for the auxiliary bands and  $\textbf{v}^{T+L}=\textbf{A}\textbf{n}^{T+L}$ is the symmetry vector for all the bands in the TETB. Lastly, $\tilde{\textbf{v}}^T$ is the regular piece of the symmetry vector $\textbf{v}^T$. For instance, the regular piece in Eq.~\eqref{vT} is given by
\begin{equation}\label{tildeV} 
    \tilde{\textbf{v}}^T = [\Gamma_2^-+\Gamma_4^-,R_4^-+R_5^+,M_1+2M_4,X_1+X_3+X_4].
\end{equation}

Besides  Eqs.~\eqref{diophantine} and \eqref{BS}, the third essential ingredient of direct solution methods is the following norm introduced in the main text
\begin{equation}\label{norm}
    ||\textbf{n}|| = \sum_{a=1}^{N_{EBR}}d_a |n_a|=\textbf{d} \cdot |\textbf{n}|, 
\end{equation}
where $\textbf{d}$ is a vector with the dimensions of the $N_{EBR}$ EBRs, and $|n_a|$ is the absolute value of the multiplicity of each EBR.
Note that $||\textbf{n}^T||=\mu^{T+L}+\mu^L=\mu^T + 2 \mu^L$, where $\mu^{T+L}=||\textbf{n}^{T+L}||$ counts the number of bands in the TETB model, and $\mu^{L}=||\textbf{n}^{L}||$ is the number of auxiliary bands for a  solution 
$\textbf{n}^T=\textbf{n}^{T+L}-\textbf{n}^L$. As the number of transverse bands $\mu^{T}$ is fixed by the symmetry vector $\textbf{v}^T$, minimizing the norm of the solution $||\textbf{n}^T||$ also minimizes the number of auxiliary bands and the complexity of the TETB model.

As previously mentioned, the main difficulty with our Diophantine equations is that they admit an infinite number of solutions. The guide to navigating this infinite sea is the norm in Eq.~\eqref{norm}, while the two direct solution approaches described below differ in how the norm is used. 

\subsubsection{Optimization Method}\label{subsec:opt}

In this method, we try to find an integer vector $\mathbf{n}^T$ with the minimal norm by solving a lattice reduction procedure with bounds \cite{aardal2000solving}.
This is equivalent to recasting the Diophantine equations as a linear optimization problem, where the norm of $\mathbf{n}^T$ is minimized subject to the constraints in Eq. \eqref{diophantine}.
Such equations can be solved using any integer linear optimization solver, and we have used the `A Mathematical Programming Language' (AMPL) \cite{fourer1987ampl} package.
This method can be summarized as follows
\begin{equation}
    \begin{aligned}
        \min \quad & ||\textbf{n}^T|| =\textbf{d} \cdot |\textbf{n}^T|\\
        \textrm{s.t.} \quad & \textbf{v}^T_{(BZ-\Gamma)}=\textbf{A}_{(BZ-\Gamma)}\textbf{n}^T \\
         & \textbf{n}^T\in\mathds{Z}^{N_{EBR}}, \\
    \end{aligned}
    \label{minimization}
\end{equation}
where the output will take the form $\{ \mathbf{n}^T, (\blacksquare)^{2T} \}$.
Ideally, the resulting vector $\mathbf{n}^T$ will have the shortest possible norm.
Note that unphysical solutions can not be used to build a viable TETB model and should be discarded, as explained in the main text and in subsection~\ref{subsec:role}.

Although the method is easy to implement using any linear optimization program, the main drawback is that it is not guaranteed to obtain the minimum norm solution in all cases. As explained in the main text, there are two reasons for this, namely, the propensity of optimization routines to get stuck around local minima and the existence of ``composite'' optimal TETB models. These problems are solved by the enumeration method, which will be considered in the following subsection.

As an example where the optimization method obtains the correct optimal solution, we consider the  PhC crystal with SG224 studied in the main text. Such a crystal is a fully connected geometry composed of dielectric rods arranged along the main diagonals of the cube. By analyzing the symmetry properties of the electromagnetic modes at the HSP, we determine the symmetry vector for the lowest six  bands of the photonic crystal
\begin{equation} \label{eq:symVec224}
    \mathbf{v}^T = [(\blacksquare)^{2T}+\Gamma_2^-+\Gamma_4^-,R_4^-+R_5^+,M_1+2M_4,X_1+X_3+X_4].
\end{equation}
Then, running the optimization method yields
\begin{equation}
   \{ A_{2u}@4b + A_{2u}@4c - A_1@2a, \;\Gamma_4^- - \Gamma_1^+\}.
\end{equation}
This solution defines a TETB model with a total of eight bands, including the six transverse bands in~\eqref{eq:symVec224}, and two auxiliary ones associated with the negative multiplicity EBR $A_1@2a$.
As two is the minimum dimension of EBRs in SG224, we know that this must be an optimal solution~\footnote{Remember that, SG224 being an unpinned group, any solution without auxiliary bands would be unphysical according to the analysis in subsection~\ref{subsec:role}.}. For details on the construction of the TETB Hamiltonian, see the main text and subsection~\ref{sec:NM_224}.

On the other hand, using the optimization method on
\begin{equation} \label{eq:c2}
    \mathbf{v}^T = [(\blacksquare)^{2T}+\Gamma_1^++\Gamma_5^+,R_1^++R_2^-+R_3^++R_3^-,2M_1+M_2,X_1+X_3+X_4].
\end{equation}
gives
\begin{equation}
   \{ A_1@2a + E@2a +T_2@2a-A_1@6d, \;\Gamma_4^- - \Gamma_1^+\}.
\end{equation}
with six auxiliary bands for a total of twelve in the TETB. This is much larger than the an the actual 
optimal solution, with only two auxiliary bands, defined by
\begin{equation}\label{c2}
    \textbf{n}^{T+L} = A_1@2a+B_2@6d\;,\;\;\; \textbf{n}^{L} = A_1@2a \;,\;\;\; (\blacksquare)^{2T} = \Gamma_4^- - \Gamma_1^+.
\end{equation}
Note that this is another example of a ``composite solution'', see also Eq.~\eqref{opt}, where the optimal TETB model for a particular crystal has at least one  EBR which is common to $\textbf{n}^{T+L}$ and $\textbf{n}^L$. 
The reason why this solution can not be obtained with the optimization method is that $A_1@2a $, being common to $\textbf{n}^{T+L}$ and $\textbf{n}^{L}$, cancels out in the solution \hbox{$\textbf{n}^{T}= \textbf{n}^{T+L} - \textbf{n}^{L}=B_2@6d$} and is invisible to any method that only computes the difference $\textbf{n}^{T}$.

\subsubsection{Enumeration Method}\label{subsec:enum}

As explained in the main text, the enumeration method was developed in an attempt to overcome the limitations 
of the linear optimization method. To avoid missing composite solutions, $\textbf{n}^{T+L}$ and $\textbf{n}^{L}$
are computed separately, first $\textbf{n}^{L}$ and then $\textbf{n}^{T+L}$. In the first step, all the possible sets of $\mu^L$ (auxiliary) bands are obtained for a specified value of~$\mu^L$. This amounts to solving a linear system

\begin{equation}\label{normA}
    ||\textbf{n}^L|| = \sum_{a=1}^{N_{EBR}}d_a n_a^L=\textbf{d} \cdot \textbf{n}^L=\mu^L, 
\end{equation}   
where the EBR multiplicities $n_a^L$ are all \textit{positive} integers. This is just a simple partition problem with a \textit{finite} number of solutions and can be solved very efficiently without requiring the use of optimization routines. For example, for SG224 there are only two solutions to~\eqref{normA} with $\mu^L=2$ bands
\begin{equation}\label{aux2}
    \{A_1@2a\;,\; A_2@2a\}, 
\end{equation}
while there are 12 with $\mu^L=4$ bands (all the EBRs in SG224 are even-dimensional)

\begin{gather}\label{aux4}
    \{A_{2u}@4c, A_{2g}@4c, A_{1u}@4c, A_{1g}@4c, A_{2u}@4b, A_{2g}@4b,\nonumber\\ A_{1u}@4b, A_{1g}@4b, E@2a, 2 A_2@2a, A_1@2a + A_2@2a, 2 A_1@2a\}, 
\end{gather}
and 28 with $\mu^L=6$,
\begin{gather}\label{aux4}
    \{B_2@6d, B_1@6d, A_2@6d, A_1@6d, T_2@2a, T_1@2a, A_2@2a + A_{2u}@4c,\nonumber\\ A_2@2a + A_{2g}@4c, A_{1u}@4c + A_2@2a, A_{1g}@4c + A_2@2a, A_2@2a + A_{2u}@4b,\nonumber\\ A_2@2a + A_{2g}@4b, A_{1u}@4b + A_2@2a, A_{1g}@4b + A_2@2a, A_2@2a + E@2a,\nonumber\\ 3 A_2@2a, A_1@2a + A_{2u}@4c, A_1@2a + A_{2g}@4c,
A_1@2a + A_{1u}@4c, A_1@2a + A_{1g}@4c,\nonumber\\A_1@2a + A_{2u}@4b, A_1@2a + A_{2g}@4b, A_1@2a + A_{1u}@4b, A_1@2a + A_{1g}@4b,\nonumber\\ A_1@2a + E@2a, A_1@2a + 2 A_2@2a, 2 A_1@2a + A_2@2a, 3 A_1@2a\}.
\end{gather}

In the second step we recast Eq.~\eqref{diophantine}, moving $\textbf{n}^{L}$ to the LHS,
\begin{equation}
    \textbf{v}^T_{(BZ-\Gamma)}+\textbf{A}_{(BZ-\Gamma)}\textbf{n}^L =\textbf{A}_{(BZ-\Gamma)}\textbf{n}^{T+L}.
\label{enum}
\end{equation}
Then we plug each of the possible sets of auxiliary bands $\textbf{n}^L$ into the LHS of Eq. (\ref{enum}) and solve for $\textbf{n}^{T+L}$. As all the EBR multiplicities $n_a^{T+L}$ must be positive, this is again a problem with only a finite number of solutions that can be efficiently solved without resorting to optimization routines. As a last step, unphysical solutions should be discarded. 

Note that the output of this method takes the form  $\{\textbf{n}^{T+L},\textbf{n}^L,(\blacksquare)^{2T}\}$ and is therefore immune to possible cancellations of common EBRs in  $\textbf{n}^T=\textbf{n}^{T+L}-\textbf{n}^L$. Also, as no optimization routines are used at any point, there is no danger of the algorithm getting stuck around local minima. Thus, we are guaranteed to obtain all the solutions with a fixed number of auxiliary bands, both normal and composite. 

For this particular approach, we have developed a code available on GitHub \cite{github} that takes as inputs the matrix $\textbf{A}$, $\textbf{v}^T$ and the number of auxiliary bands $\mu^L$.
The code automatically checks all the solutions and discards unphysical ones.
In order to find the optimal solutions, we just have to use it for increasing numbers of auxiliary bands ($\mu^L=0,1,\ldots$) until the code outputs the first physical solutions, which, by construction, are necessarily optimal. Of course, if for any reason we are interested in solutions with more auxiliary bands than strictly necessary, we can use the code to generate solutions for higher values of $\mu^L$. For the rest of this subsection, we will give a few examples for SG224.

The first example is meant to illustrate the possibility of multiple optimal solutions and is based on the symmetry vector
\begin{equation} 
    \mathbf{v}^T = [(\blacksquare)^{2T}+\Gamma_1^++\Gamma_5^+,R_4^-+R_5^+,M_1+M_3+M_4,X_1+X_3+X_4],
\end{equation}
for which the code output includes three optimal solutions 
\begin{gather} \label{solsc1}
\{A_{1g}@4c + A_{2u}@4c,\, A_1@2a,\, \Gamma_4^- - \Gamma_1^+\},\nonumber\\
\{A_{1g}@4b + A_{2u}@4b,\, A1@2a, \,\Gamma_4^- - \Gamma_1^+\}\;\;,\;\; \{A_1@2a + T_2@2a,\, A_1@2a, \, \Gamma_4^- - \Gamma_1^+\}.
\end{gather}
Having several solutions to choose from allows the user to select the one that best suits each particular problem, making the construction of the TETB even more convenient. Note that the third solution is composite.

So far, all our examples for SG224 have involved the same surrogate representation $(\blacksquare)^{2T} = \Gamma_4^- - \Gamma_1^+$, which agrees with the particular solution of the compatibility relations listed in Ref. \cite{christensen2022location}. However, if we consider the symmetry vector
\begin{equation} \label{eq:c14}
    \mathbf{v}^T = [(\blacksquare)^{2T}+\Gamma_2^++\Gamma_4^+,R_4^++R_5^-,M_2+2M_4,X_2+X_3+X_4],
\end{equation}
the code finds the following optimal solution
\begin{equation}
\{A_{2g}@4b+A_{2g}@4c,\,A_2@2a,\,\Gamma_4^+ - \Gamma_1^-\},
\end{equation}
with two auxiliary bands and a surrogate representation that the reader may recognize as the one appropriate for a magnetic formulation, i.e., as the difference between the axial vector and pseudoscalar representations. 
As we are using a $\Gamma$-agnostic method, this result is automatically found by the code. Note that this surrogate representation satisfies Eq.~\eqref{surro} for $(l,p,q)=(1,1,0)$. 

If we insisted in using the more conventional surrogate representation $(\blacksquare)^{2T} = \Gamma_4^- - \Gamma_1^+$, then the ``optimal" solution would be given by
\begin{equation}
\{B_1@12f,\,A_1@6d,\,\Gamma_4^- - \Gamma_1^+\},
\end{equation}
that involves six auxiliary bands instead of two and would lead to a much more cumbersome TETB model. In this example, the form of the alternative surrogate representation was relatively easy to guess. To see that this is not always the case, consider the symmetry vector
\begin{equation} \label{eq:c6}
    \mathbf{v}^T = [(\blacksquare)^{2T}+\Gamma_1^++\Gamma_2^++\Gamma_3^-,R_1^++R_3^-+R_5^-,M_1+M_2+M_4,X_1+X_2+X_4],
\end{equation}
for which the optimal solution is given by
\begin{equation}
\{A_1@2a+A_2@2a+E_u@4b,\,A_{1g}@4c+A_2@2a,\,-\Gamma_1^+ + \Gamma_2^--\Gamma_2^++\Gamma_4^-+\Gamma_5^--\Gamma_5^+\}.
\end{equation}
Thus, the optimal solution in this example, besides being composite, involves a very ``unconventional'' surrogate
representation. Note that this surrogate representation satisfies Eq.~\eqref{surro} for $(l,p,q)=(-1,0,0)$.

We finish this subsection by pointing out that, for the 68 space groups in Table~\ref{tab:pinnedSGs}, the code will return optimal solutions without auxiliary bands if we set $\mu^L=0$ or, for higher values of $\mu^L$, with the appropriate auxiliary bands to build a TETB model compatible with a $\mathbf{k}\cdot\mathbf{p}$ description of the long wavelength limit of the modes at vanishing frequencies, as explained in subsection~\ref{subsec:role}.

\subsection{Real-space Invariants (RSI) based method} \label{subsec:RSI}

As an alternative approach to solving Eq. \eqref{matrix_eq}, we can adapt the theory of Real Space Invariants (RSIs) introduced in Refs. \cite{song2020twisted,xu2021filling}.
At each WP, RSIs are invariants computed from the multiplicities $\mathbf{m}$ of the occupied site symmetry group representations or Wannier orbitals.
In particular, the RSIs parameterize whether or not two configurations of occupied orbitals can be adiabatically deformed into each other without breaking the system's symmetries.
All RSIs take integer values for topologically trivial gapped bands and symmetry-indicated fragile topological bands.
Finally, for gapless or topologically nontrivial bands, the RSIs take rational values \cite{hwang2019fragile,song2020twisted,xu2021three}.
Generally, the number of  RSIs at maximal WPs in a space group will be smaller than the number of EBRs.
Additionally, as discussed below, a subset (or linear combinations) of the RSIs of every space group can be computed from the symmetry vector in momentum space. 

Below, we combine the position and momentum space expressions for the RSIs to solve Eq.~\eqref{matrix_eq}.
The solution procedure is as follows:
We will first complete the symmetry vector in momentum space at the $\Gamma$ point $(\blacksquare)^{2T}$, based on the result in Ref. \cite{christensen2022location}. 
Next, we compute all RSIs that can be determined from the symmetry vector, including enough photonic bands to ensure we get integers for all RSIs. 
Since there are fewer RSIs than EBRs, this gives us a smaller set of Diophantine equations, which, in simple cases, can even be solved by hand for the occupied site symmetry group representations at each WP. On the other hand, the fact that we are considering a linear system of lower rank than the original implies that some of the solutions obtained by this method may not solve 
Eq. \eqref{matrix_eq}, and this should be checked at the end of the calculation.

To see how this works, we will first analyze the RSIs of photonic crystals in SG $P2$ (No. 3) in Sec. \ref{subsec:p2rsis} as a simple example of the procedure.
Then, in Sec. \ref{subsec:224rsis}, we will apply this machinery to solve Eq. \eqref{matrix_eq} for the photonic crystal with SG No. 224 ($Pn\Bar{3}m$).

\subsubsection{Photonic RSIs in Space Group $P2$} \label{subsec:p2rsis}

We first consider SG $P2$ (No. 3), the simplest nontrivial space group, for a clear example of RSIs.
SG $P2$ is generated by a twofold rotation, $C_{2y}: (x,y,z) \to (x,-y,z)$, and three translations by primitive lattice vectors which we take for simplicity to be $\mathbf{t}_1=\hat{x}$, $\mathbf{t}_2=\hat{y}$, and $\mathbf{t}_3=\hat{z}$.
SG $P2$ has four maximal WPs denoted by $\mathbf{q}=1a$, $1b$, $1c$, and $1d$,
with site-symmetry groups generated by  $g_{1a}=\{C_{2y}|\mathbf{0}\}$, $g_{1b}=\{C_{2y}|\mathbf{t}_3\}$, $g_{1c}=\{C_{2y}|\mathbf{t}_1\}$, and $g_{1d}=\{C_{2y}|\mathbf{t}_1 + \mathbf{t}_3\}$, respectively.
There is also a generic WP $2e$ with a trivial site-symmetry group.
Note that the representative positions of WPs are $1a: (0,y,0)$, $1b: (0,y,1/2)$, $1c: (1/2,y,0)$, $1d: (1/2,y,1/2)$, and $2e: \{(x,y,z),(-x,y,-z)\}$, where $x,y,z$ are free parameters.
This indicates that changing the position of the orbital through those parameters preserves the symmetries imposed by the SG and the system's topology.
The coordinates and site symmetry groups for these Wyckoff positions are summarized in Table \ref{tab:p2wyckoff}.

{
\renewcommand{\arraystretch}{1.15}
\begin{table}
    \centering
    \begin{tabular}{c|cc}
        \hline \hline
        WP ($\mathbf{q}$) & Coordinates & $G_\mathbf{q}$\\
        \hline
        $1a$ & $(0,y,0)$ & $\{C_{2y}|\mathbf{0}\}$ \\
        $1b$ & $(0,y,1/2)$ & $\{C_{2y}|\mathbf{t}_3\}$ \\
        $1c$ & $(1/2,y,0)$ & $\{C_{2y}|\mathbf{t}_1\}$ \\
        $1d$ & $(1/2,y,1/2)$ & $\{C_{2y}|\mathbf{t}_1 + \mathbf{t}_3\}$ \\
        \multirow{2}{*}{$2e$} & $(x,y,z)$, & \multirow{2}{*}{$\{E|\mathbf{0}\}$} \\
        & $(-x,y,-z)$ & \\
        \hline \hline
    \end{tabular}
    \hspace{5mm}
    \begin{tabular}{c|cc}
        \hline \hline
        Irrep ($\rho$) & $1$ & $C_2$\\
        \hline
        $A$ & $1$ & $1$ \\
        $B$ & $1$ & $-1$ \\
        \hline \hline
    \end{tabular}
    \caption{Left: Wyckoff positions (WPs) of SG $P2$.
    $G_\mathbf{q}$ denotes site-symmetry group of WP $\mathbf{q}$.
    Right: Irreducible representations $\rho$ of point group $C_2$.
    }
    \label{tab:p2wyckoff}
\end{table}
}

In this SG, any orbital $\rho_\mathbf{q}$ localized at a maximal WP $\textbf{q}$ ($\mathbf{q}=1a$, $1b$, $1c$, $1d$) transforms as either the $\rho= A$ or $B$ irrep of the point group $C_2$, which is isomorphic to the site symmetry group $G_{\textbf{q}}$ (See Table \ref{tab:p2wyckoff}).
$A$ has eigenvalue $+1$ while $B$ has $-1$ with respect to $C_2$. 
We denote the multiplicity of each orbital as $m(\rho@\mathbf{q})$.
Any band representation (BR) corresponding to a topologically trivial band can be induced from a set of orbitals located at maximal WPs. 
Consequently, one can assign a set of multiplicities corresponding to such orbitals for a given trivial BR.
We will denote this set as $\{m(\rho_\mathbf{q})\}$.

However, as we will see, $\{m(\rho@\mathbf{q})\}$ may not be uniquely determined since each $m(\rho@\mathbf{q})$ is not invariant under adiabatic deformations that preserve the space group symmetry.
Nevertheless, some linear combinations of $m(\rho@\mathbf{q})$-s may remain invariant under such transformations.
In the study of crystalline materials, these invariant linear combinations are known as RSIs.

We will now show how to derive the RSIs in SG $P2$.
We know that a single orbital at one of the maximal WP transforming in either the $A$ or $B$ representation of $G_\mathbf{q}$ cannot be moved adiabatically to a different WP; only the $y$ coordinate of the maximal WP is a free parameter.
On the contrary, if we have two orbitals, one $A@1a$ and one $B@1a$, they can be hybridized into a pair of $C_{2y}$-related orbitals.
For clear illustration, let us denote the two orbitals transforming as $A@1a$ and $B@1a$ as $\phi_{A@1a}$ and $\phi_{B@1a}$ respectively.
Then, their linear combinations $\Phi_\pm = \frac{1}{\sqrt{2}} (\phi_{A@1a} \pm \phi_{B@1a})$ are related by $C_{2y}$ to each other.
In this case, $\Phi_+$ and $\Phi_-$ can move to $(x,y,z)$ and $(-x,y,-z)$ respectively because their configuration respects $C_{2y}$ for any value of free parameters $(x, y,z)$, which characterize the generic WP $2e$.
Hence, these hybridized orbitals $\Phi_\pm$ can be adiabatically moved to the WP $2e$ and become $A@2e$ while respecting the symmetry properties and topology of the band representation. 

Conversely, an $A@2e$ orbital can be deformed into a pair of $A@1a + B@1a$ orbitals.
Hence, in the adiabatic processes, neither $m(A@1a)$ nor $m(B@1a)$ are invariant.
However, the difference $-m(A@1a)+m(B@1a)$ remains invariant.
Since no adiabatic process can change $-m(A@1a) + m(B@1a)$, it must be a topological invariant in SG $P2$.
We thus identify it with a real space invariant, which we call $\delta_{1a}$.

Repeating the preceding analysis for the other maximal WP allows us to define the RSIs for space group $P2$ as

\begin{gather}
    \delta_{1a} = -m(A@1a)+m(B@1a), \nonumber \\
    \delta_{1b} = -m(A@1b)+m(B@1b), \nonumber \\
    \delta_{1c} = -m(A@1c)+m(B@1c), \nonumber \\
    \delta_{1d} = -m(A@1d)+m(B@1d).
    \label{eq:RSI_SG2_real}
\end{gather}

From this definition, we can easily deduce which WPs must be occupied by Wannier orbitals based on nonzero RSIs.
By construction, when it is possible to move every orbital at a WP $\mathbf{q}$ to other WP(s), $\delta_{\mathbf{q}}$ is zero.
Thus, if an RSI $\delta_{\mathbf{q}}$ is nonzero, the relevant WP $\mathbf{q}$ must be occupied by at least one orbital.
In other words, a nonzero RSI indicates the existence of symmetry-pinned orbital(s) at the corresponding WP.

The above discussion about adiabatic deformations can be translated into subgroup-group and induction-subduction relations.
We refer the reader to Ref. \cite{xu2021three} for a general discussion on RSIs.
The RSIs for every magnetic and nonmagnetic SGs have been tabulated in the past and are enumerated in Ref. \cite{xu2021three} and the BCS \cite{aroyo2011crystallography}. 

The RSIs in Eq. \eqref{eq:RSI_SG2_real} are defined in terms of orbital multiplicities localized in real space. 
RSIs formulated in this way are called the local RSIs \cite{xu2021three}.
In the case of SG $P2$, all four local RSIs can be completely determined by the multiplicities of the momentum-space irreps, i.e., in terms of the symmetry vector $\mathbf{v}$ \cite{van2018higher,hwang2019fragile,song2020twisted,xu2021three,aroyo2011crystallography}.

To construct a mapping between the RSIs and multiplicities of Wannier orbitals, we define $\boldsymbol{\delta}=(\delta_{1a},\delta_{1b},\delta_{1c},\delta_{1d})^T$ and

\begin{align} \label{eq:orb_multi}
    \mathbf{m}=[m(A@1a), m(B@1a), m(A@1b), m(B@1b), m(A@1c), m(B@1c), m(A@1d), m(B@1d)]^T.
\end{align}

Then, Eq. \eqref{eq:RSI_SG2_real} can be rewritten as

\begin{align}
    \boldsymbol{\delta}
    = \begin{pmatrix}
    -1 & 1 & 0 & 0 & 0 & 0 & 0 & 0 \\
    0 & 0 & -1 & 1 & 0 & 0 & 0 & 0 \\
    0 & 0 & 0 & 0 & -1 & 1 & 0 & 0 \\
    0 & 0 & 0 & 0 & 0 & 0 & -1 & 1
    \end{pmatrix}
    \mathbf{m}.
    \label{eq:lRSI_SG2}
\end{align}

A set of Wannier orbitals with multiplicities $\mathbf{m}$ induces a BR with symmetry vector $\mathbf{v}$.
In Eq. \eqref{eq:orb_multi}, $\mathbf{m}$ counts the multiplicities of all types of Wannier orbitals at maximal WPs.
Note that $\mathbf{m}$ coincides with $\mathbf{n}^T$ defined in Eq. \eqref{matrix_eq} except in exceptional cases defined in Ref. \cite{bradlyn2017topological} (In such cases, $\mathbf{n}$ contains only site-symmetry group representations at a subset of maximal WPs that is sufficient to generate all EBRs.
In most cases, this coincides with the set of all maximal WPs).
Because a Wannier orbital at a maximal WP induces an EBR, $\mathbf{v}$ can be computed by the EBR matrix $\mathbf{A}$ in $P2$ such that $\mathbf{v} = \mathbf{A} \textbf{m}$.
Thus, for a given symmetry vector $\mathbf{v}$, $\mathbf{m}$ can be solved by Eq. \eqref{pseudoinverse}.
Explicitly, Eq. \eqref{pseudoinverse} yields

\begin{align}\label{eq:orbital_irrep_SG2}
    m(A@1a) &= \frac{1}{2}(-n_{\Gamma_1}+n_{A_1}+n_{B_1}+n_{C_1})-z_4-z_6-z_8, \nonumber \\
    m(A@1b) &= \frac{1}{2}(n_{\Gamma_1}-n_{A_1}-n_{B_1}+n_{C_1}) + z_4, \nonumber \\
    m(A@1c) &= \frac{1}{2}(n_{\Gamma_1}-n_{A_1}+n_{B_1}-n_{C_1})+z_6, \nonumber \\
    m(A@1d) &= \frac{1}{2}(n_{\Gamma_1}+n_{A_1}-n_{B_1}-n_{C_1}) + z_8, \nonumber \\
    m(B@1a) &= n_{\Gamma_2}-z_4-z_6-z_8, \nonumber \\
    m(B@1b) &= z_4, \quad
    m(B@1c) = z_6, \quad
    m(B@1d) = z_8.
\end{align}
Here, $\Gamma$, $A$, $B$ and $C$ refer to the HSPs of SG $P2$, and $\Gamma_{1,2}$, $A_{1,2}$, $B_{1,2}$, and $C_{1,2}$ refer to little group irreps in the BCS notation.
Also, $n_{\rho_\mathbf{k}}$ denotes the multiplicity of the irrep $\rho_\mathbf{k}$, and $z_{4,6,8}$ are free integer parameters defined in Eq. \eqref{pseudoinverse}.
Substituting Eq. \eqref{eq:orbital_irrep_SG2} into Eq. \eqref{eq:lRSI_SG2}, we obtain

\begin{align} \label{eq:RSI_SG2_momentum}
    & \delta_{1a} = n_{\Gamma_2} + \frac{1}{2} \left( n_{\Gamma_1} - n_{A_1} - n_{B_1} - n_{C_1} \right), \nonumber \\
    & \delta_{1b} = \frac{1}{2} \left( -n_{\Gamma_1} + n_{A_1} + n_{B_1} - n_{C_1} \right), \nonumber \\
    & \delta_{1c} = \frac{1}{2} \left( - n_{\Gamma_1} + n_{A_1} - n_{B_1} + n_{C_1} \right), \nonumber \\
    & \delta_{1d} = \frac{1}{2} \left( - n_{\Gamma_1} - n_{A_1} + n_{B_1} + n_{C_1} \right).
\end{align}

Let us emphasize that Eq. \eqref{eq:RSI_SG2_momentum} is uniquely determined from the symmetry data, while each multiplicity of Wannier orbital in Eq. \eqref{eq:orbital_irrep_SG2} depends on free parameters $z_{4,6,8}$. 
For a general space group, some local RSIs may not be determined uniquely from the momentum space symmetry vector.
In these cases, linear combinations of RSIs can be constructed that are determined uniquely from the symmetry vector;
these linear combinations are known as composite RSIs, and they count the multiplicities of orbitals at multiple WPs.
Composite RSIs are tabulated in Ref. \cite{xu2021three}.
When local RSIs cannot be determined from symmetry vectors, composite RSIs can be computed instead.

In the previous discussion, we introduced RSIs in general terms for a simple SG.
Our objective is to compute them to derive a set of pseudo-orbitals that yield a given symmetry vector $\mathbf{v}^T$.
To compute the RSIs from a given symmetry vector $\mathbf{v}^T$ of a 3D photonic band structure, some additional information on $(\blacksquare)^{2T}$ is necessary.
Maxwell's equations and symmetry-compatibility relations along high-symmetry lines and planes impose strong conditions on $(\blacksquare)^{2T}$ \cite{watanabe2018space,christensen2022location}.
Transverse ($2T$) modes must be expressible as the difference between the vector and the trivial representations of the little group of $\Gamma$~\footnote{This applies to a formulation in terms of electric fields. In a magnetic formulation, we should consider the difference between the axial vector and pseudoscalar representations. Note, however, that vectors and axial vectors transform the same way under SG3, and there is no difference between the electric and magnetic formulations}.
As a result, $(\blacksquare)^{2T}$ can be expressed as a linear combination of irreps $\rho_{i[\Gamma]}$, i.e. $(\blacksquare)^{2T}= \oplus_i n_{i[\Gamma]} \rho_{i[\Gamma]}$, where $n_{i[\Gamma]}$ can be a negative integer.
The particular expressions of $(\blacksquare)^{2T}$ depend on the given SG, and this information is listed in Ref. \cite{christensen2022location} for the 230 SGs.
RSIs were originally implemented for ``regular'' irrep multiplicities ($n_{i[\Gamma]} \ge 0$).
Nevertheless, although $(\blacksquare)^{2T}$ can be irregular, meaning that some irreps can have negative multiplicities ($n_{i[\Gamma]}<0$), RSIs can still be computed.
This is so because the RSIs are defined by a (linear) equation between some matrix and $n_{i[\mathbf{k}]}$.

Let us explain how the RSI analysis can efficiently solve the Diophantine equation in Eq. \eqref{diophantine}.
For this, we consider two possible examples of $\mathbf{v}^T$ in SG $P2$;
$\mathbf{v}^{(1)} = [(\blacksquare)^{2T}, 2A_1, B_1 + B_2, 2C_1, D_1 + D_2, 2E_1, 2Y_1, 2Z_2]$ and $\mathbf{v}^{(2)} = [(\blacksquare)^{2T}, 2A_1, 2B_1, 2C_1, 2D_1, 2E_1, 2Y_1, 2Z_2]$.
Both have the dimension ${\rm dim}(\mathbf{v}^T) = 2$. 
First, note that $(\blacksquare)^{(2T)}=2\Gamma_2$ in SG $P2$ \cite{watanabe2018space,christensen2022location}. 
Then, by using Eq. \eqref{eq:RSI_SG2_momentum}, the RSI indices $\boldsymbol{\delta} = (\delta_{1a},\delta_{1b},\delta_{1c},\delta_{1d})$ can be computed for $\mathbf{v}^{(1,2)}$ simply following Eq. \eqref{eq:RSI_SG2_real}.
In the case of $\mathbf{v}^{(1)}$, the corresponding RSI indices are $\boldsymbol{\delta}^{(1)}=(-1/2,1/2,3/2,1/2)$.
When at least one RSI in $\boldsymbol{\delta}$ is fractional, i.e., $\boldsymbol{\delta}^{(1)} \notin \mathbb{Z}^4$, $\mathbf{v}^{(1)}$ has stable topology with nontrivial symmetry indicator and does not allow integer-valued $\mathbf{n}$ or $m(\rho_\mathbf{q})$.
Thus, $\mathbf{v}^{(1)}$ is not Wannierizable.
Instead, one can consider a bigger set including higher energy bands with symmetry vector $\mathbf{v}'$ such that $\mathbf{v}^{(1)}+\mathbf{v}'$ corresponds to integer-valued RSI indices $\boldsymbol \delta$ and allows the Wannierzation.

On the other hand, $\mathbf{v}^{(2)}$ has an integer-valued RSIs with $\mathbf{\delta}^{(2)}=(-1,1,1,1)$.
This implies

\begin{gather}
    -m(A@1a)+m(B@1a)=-1, \nonumber \\
    -m(A@1b)+m(B@1b)=1, \nonumber \\
    -m(A@1c)+m(B@1c)=1 \nonumber \\
    -m(A@1d)+m(B@1d)=1.
\end{gather}

To solve these equations, we assume that all the orbitals are localized at maximal WPs $\mathbf{q}=1a, 1b, 1c, 1d$ because orbitals at any non-maximal WP can always be moved to $\mathbf{q}$.
With this assumption, ${\rm dim}(\mathbf{v}^{(2)})=2$ imposes $\sum_{\mathbf{q}} \sum_{\rho=A,B} m(\rho_\mathbf{q}) = 2$.
One can easily find a particular solution, $m(A@1a)=m(B@1b)=m(B@1c)=1$, $m(A@1d)=-1$, which is also a solution to Eq. \eqref{matrix_eq}, and thus we express $\mathbf{v}^{(2)}$ as $\mathbf{v}^{(2)}=\mathbf{A}\mathbf{n}^{T}$

\begin{align}
    \label{eq:RSI_resulst_SG2}
    \mathbf{n}^{T} = A@1a + B@1b + B@1c - A@1d.
\end{align}

This implies that one can construct a TB model for the photonic bands corresponding to $\mathbf{v}^{(2)}$ by introducing an auxiliary pseudo-orbital $A$ at WP $1d$. 
Note that, to regularize the polarization singularity at $\Gamma$ by adding auxiliary band(s), the auxiliary band must span the trivial irrep \cite{christensen2022location}.
Consistent with this condition, $A@1d$ orbital induces the EBR $A@1d$ with the trivial irrep $\Gamma_1$.

\subsubsection{Photonic RSIs in Space Group $Pn\bar{3}m$} \label{subsec:224rsis}

Let us consider, again, the crystal with SG $Pn\bar{3}m$ (No. 224).
In this SG, there are 5 maximal WPs: $2a: (1/4,1/4,1/4)$ with site-symmetry group $G_{2a}=\bar{4}3m$, $4b: (0,0,0)$ with $G_{4b}=\bar{3}m$, $4c: (1/2,1/2,1/2)$ with $G_{4c}=\bar{3}m$, $6d: (1/4,3/4,3/4)$ with $G_{6d}=\bar{4}2m$, and $12f: (1/2,1/4,3/4)$ with $G_{12f}=222$.
For these maximal WPs, the RSIs are \cite{xu2021three}

\begin{align} \label{eq:RSI_SG224_real}
    & \delta_{a} = m(A_1)-m(A_2)-m(T_2)+m(T_1), \nonumber \\
    & \delta_{d} = -m(A_1) + m(B_1) + m(B_2) - m(A_2), \nonumber \\
    & \delta_{f} = m(A_1) + m(B_1) - m(B_3) - m(B_2) \quad ({\rm mod} \, 2), \nonumber \\
    & \delta_{\mathbf{q},1} = m(A_{1g}) - m(A_{1u}) + m(A_{2g}) - m(A_{2u}), \nonumber \\
    & \delta_{\mathbf{q},2} = -m(E_g) + m(E_u),
\end{align}
where $\mathbf{q}=b,c$
(We omit the multiplicities of WPs for simpler notation.).
RSIs can also be defined for nonmaximal WPs ($8e,12g,24h,24i,24j,24k$).

To avoid complications, let us focus on the RSIs defined in Eq. \eqref{eq:RSI_SG224_real} for the maximal WPs $a,b,c,d,f$ because any orbitals at nonmaximal WPs can be adiabatically moved to the maximal WPs
\footnote{Depending on the SG, the symmetry vector may not determine the full set of local RSIs defined at maximal WPs.
In this case, the composite RSIs can be computed, indicating the WPs where orbitals are occupied.
Then, the Diophantine equation in Eq. \eqref{matrix_eq} can be solved for those occupied WPs according to the definition of composite RSIs.}.
We denote those RSIs as $\tilde{\boldsymbol{\delta}} = (\delta_a, \delta_{b,1}, \delta_{b,2}, \delta_{c,1}, \delta_{c,2}, \delta_d, \delta_f)$ collectively.
This reduced set of RSIs $\tilde{\boldsymbol{\delta}}$ can be uniquely computed from a given symmetry vector $\mathbf{v}^T$, and the mapping from $\mathbf{v}^T$ to $\tilde{\boldsymbol{\delta}}$ is accessible via the BCS \cite{aroyo2011crystallography}.

As we did in Sec. \ref{subsec:numerical}, from the analysis of the symmetry properties of the Bloch electromagnetic modes supported by the PhC, we obtain the following symmetry vector

\begin{align} \label{eq:sv224}
    \mathbf{v}^T = [(\blacksquare)^{2T}+\Gamma_2^-+\Gamma_4^-,R_4^-+R_5^+,M_1+2M_4,X_1+X_3+X_4].
\end{align}
which is representative of the lowest six active bands in the photonic spectrum.
This is consistent with the symmetry-constrained irrep content $\Gamma$ proposed by Ref. \cite{christensen2022location}, $(\blacksquare)^{2T} = \Gamma_4^- - \Gamma_1^+$.
For $\mathbf{v}^T$ the RSI indices are $\tilde{\boldsymbol{\delta}}=(-1,-1,0,-1,0,0,0)$.
Thus, only the three RSIs related to WP $a,b,c$ are nonzero, i.e. $\delta_{a} = \delta_{b,1} = \delta_{c,1} =-1$.
To solve the RSI equations (Eq. \eqref{eq:RSI_SG224_real}), we further use ${\rm dim}(\mathbf{v}^T)=6$ and the irreps at $\Gamma$, which are $-\Gamma_1^+ + \Gamma_2^- + 2\Gamma_4^-$.

Eqs. \eqref{eq:RSI_SG224_real} admit infinitely many solutions, but we are interested only in those with the minimum number of auxiliary bands. On the other hand, we know from the analysis in subsection \ref{subsec:role} that the auxiliary bands should cancel the negative irrep $\Gamma_1^+$ in $(\blacksquare)^{2T}$. A look at \textit{BandRep} in the BCS shows that the lowest dimensional  EBR that contains $\Gamma_1^+$ is $A_1@2a$, which has dimension two. Thus, by placing just one $A_1$-orbital at the WP $1a$ we satisfy the condition  $\delta_{a}=-1$. In order get $\delta_{b,1}=-1$ we must take  $m(A_{1u}@4b)=1$ \textit{or} $m(A_{2u}@4b)=1$. Similarly, we can get $\delta_{c,1}=-1$ by taking $m(A_{1u}@4c)=1$ or $m(A_{2u}@4c)=1$. There are thus four possible combinations that satisfy the RSI equations~\eqref{eq:RSI_SG224_real} with only two auxiliary bands, namely
\begin{equation}\label{4sols}
\begin{aligned}
    \textbf{n}^{T} &= A_{1u}@4b+A_{1u}@4c-A_1@2a\\
    \textbf{n}^{T} &= A_{1u}@4b+A_{2u}@4c-A_1@2a\\
    \textbf{n}^{T} &= A_{2u}@4b+A_{1u}@4c-A_1@2a\\
    \textbf{n}^{T} &= A_{2u}@4b+A_{2u}@4c-A_1@2a
    \end{aligned}
\end{equation}
Now,  the rank of the original diophantine equations~\eqref{matrix_eq} for SG224 is thirteen, while there are only seven equations in~\eqref{eq:RSI_SG224_real}. Thus, not all solutions to the RSI constraints are guaranteed to satisfy the original equations. Indeed, computing the symmetry vectors for the four solutions in~\eqref{4sols} shows that only the last one reproduces  Eq.~\eqref{eq:sv224}. Thus we take
\begin{align}
    \label{eq:RSI_resulst_SG224}
    \mathbf{n}^T = A_{2u}@{4b}  +  A_{2u}@{4c}  -  A_1@{2a},
\end{align}
which agrees with the solution obtained by the two other methods.

\section{Examples of TETB hamiltonians} \label{sec:SM_221}

\subsection{TETB Hamiltonian for SG224} \label{sec:NM_224}

\subsubsection{Diophantine equation for SG224}

In this section, we explain how to represent the Diophantine equation in such a way that its solution can be easily implemented on a computer.

With that objective in mind, let us create a vector with the length of the number of irreps of the space group without including the $\Gamma$ point where the entries correspond to the multiplicity of that irrep. For example, in the particular case considered

\begin{equation}
    \mathbf{v}^T_{(BZ-\Gamma)} = [R_4^-+R_5^+,M_1+2M_4,X_1+X_3+X_4].
\end{equation}

The numerical vector associated with it will be

\begin{equation}
    \mathbf{v}^T_{(BZ-\Gamma)} = (0,0,0,0,0,0,0,1,1,0,1,0,0,2,1,0,1,1),
\end{equation}
where we have followed the order

\begin{align}
\{R_1^+,R_1^-,R_2^+,R_2^-,R_3^+,R_3^-,R_4^+,R_4^-,R_5^+,R_5^-,M_1,M_2,M_3,M_4,X_1,X_2,X_3,X_4\},
\end{align}
for the classification of irreps in SG No. 224.

We will now construct matrix $\textbf{A}_{(BZ-\Gamma)}$ following the same strategy. Column by column, we can follow the same recipe to determine the numerical symmetry vector of each EBR. For example, for EBR $A_1@2a$, the symmetry vector is

\begin{equation}
    \mathbf{v}^T_{(BZ-\Gamma)}(A_1@2a) = [R_1^+ + R_2^+, M_1, X_1] = (1,0,1,0,0,0,0,0,0,0,1,0,0,0,1,0,0,0).
\end{equation}

Following the same procedure for the 25 EBRs of SG $Pn\bar{3}m$ (No. 224), one can build $\textbf{A}_{(BZ-\Gamma)}$, getting:

\setcounter{MaxMatrixCols}{30}
\begin{equation}
    A_{(BZ-\Gamma)} =
    \begin{pNiceMatrix}[first-row,last-col]
        \rb{A_1@2a} & \rb{A_2@2a} & \rb{E@2a} & \rb{T_1@2a} & \rb{T_2@2a} & \rb{A_{1g}@4b} & \rb{A_{1u}@4b} & \rb{A_{2g}@4b} & \rb{A_{2u}@4b} & \rb{E_{g}@4b} & \rb{E_u@4b} & \rb{A_{1g}@4c} & \rb{A_{1u}@4c} & \rb{A_{2g}@4c} & \rb{A_{2u}@4c} & \rb{E_g@4c} & \rb{E_u@4c} & \rb{A_1@6d} & \rb{A_2@6d} & \rb{B_1@6d} & \rb{B_2@6d} & \rb{A_1@12f} & \rb{B_1@12f} & \rb{B_2@12f} & \rb{B_3@12f} &  \\
        1 & 0 & 0 & 0 & 0 & 1 & 0 & 0 & 0 & 0 & 0 & 0 & 0 & 0 & 1 & 0 & 0 & 0 & 0 & 0 & 1 & 0 & 0 & 1 & 0 & R_1^+ \\
        0 & 1 & 0 & 0 & 0 & 0 & 1 & 0 & 0 & 0 & 0 & 0 & 0 & 1 & 0 & 0 & 0 & 0 & 1 & 0 & 0 & 0 & 0 & 1 & 0 & R_1^- \\
        0 & 1 & 0 & 0 & 0 & 0 & 0 & 1 & 0 & 0 & 0 & 0 & 1 & 0 & 0 & 0 & 0 & 0 & 1 & 0 & 0 & 0 & 0 & 0 & 1 & R_2^+ \\
        1 & 0 & 0 & 0 & 0 & 0 & 0 & 0 & 1 & 0 & 0 & 1 & 0 & 0 & 0 & 0 & 0 & 0 & 0 & 0 & 1 & 0 & 0 & 0 & 1 & R_2^- \\
        0 & 0 & 1 & 0 & 0 & 0 & 0 & 0 & 0 & 1 & 0 & 0 & 0 & 0 & 0 & 0 & 1 & 0 & 1 & 0 & 1 & 0 & 0 & 1 & 1 & R_3^+ \\
        0 & 0 & 1 & 0 & 0 & 0 & 0 & 0 & 0 & 0 & 1 & 0 & 0 & 0 & 0 & 1 & 0 & 0 & 1 & 0 & 1 & 0 & 0 & 1 & 1 & R_3^- \\
        0 & 0 & 0 & 1 & 0 & 0 & 0 & 1 & 0 & 1 & 0 & 0 & 1 & 0 & 0 & 0 & 1 & 0 & 0 & 1 & 0 & 1 & 1 & 0 & 1 & R_4^+ \\
        0 & 0 & 0 & 0 & 1 & 0 & 0 & 0 & 1 & 0 & 1 & 1 & 0 & 0 & 0 & 1 & 0 & 1 & 0 & 0 & 0 & 1 & 1 & 0 & 1 & R_4^- \\
        0 & 0 & 0 & 0 & 1 & 1 & 0 & 0 & 0 & 1 & 0 & 0 & 0 & 0 & 1 & 0 & 1 & 1 & 0 & 0 & 0 & 1 & 1 & 1 & 0 & R_5^+ \\
        0 & 0 & 0 & 1 & 0 & 0 & 1 & 0 & 0 & 0 & 1 & 0 & 0 & 1 & 0 & 1 & 0 & 0 & 0 & 1 & 0 & 1 & 1 & 1 & 0 & R_5^- \\
        1 & 0 & 1 & 0 & 1 & 1 & 0 & 0 & 1 & 1 & 1 & 1 & 0 & 0 & 1 & 1 & 1 & 1 & 1 & 0 & 2 & 1 & 1 & 2 & 2 & M_1 \\
        0 & 1 & 1 & 1 & 0 & 0 & 1 & 1 & 0 & 1 & 1 & 0 & 1 & 1 & 0 & 1 & 1 & 0 & 2 & 1 & 1 & 1 & 1 & 2 & 2 & M_2 \\
        0 & 0 & 0 & 1 & 1 & 1 & 1 & 0 & 0 & 1 & 1 & 1 & 1 & 0 & 0 & 1 & 1 & 1 & 0 & 1 & 0 & 3 & 1 & 1 & 1 & M_3 \\
        0 & 0 & 0 & 1 & 1 & 0 & 0 & 1 & 1 & 1 & 1 & 0 & 0 & 1 & 1 & 1 & 1 & 1 & 0 & 1 & 0 & 1 & 3 & 1 & 1 & M_4 \\
        1 & 0 & 1 & 0 & 1 & 1 & 0 & 0 & 1 & 1 & 1 & 1 & 0 & 0 & 1 & 1 & 1 & 2 & 0 & 1 & 1 & 2 & 2 & 1 & 1 & X_1 \\
        0 & 1 & 1 & 1 & 0 & 0 & 1 & 1 & 0 & 1 & 1 & 0 & 1 & 1 & 0 & 1 & 1 & 1 & 1 & 2 & 0 & 2 & 2 & 1 & 1 & X_2 \\
        0 & 0 & 0 & 1 & 1 & 0 & 0 & 1 & 1 & 1 & 1 & 1 & 1 & 0 & 0 & 1 & 1 & 0 & 1 & 0 & 1 & 1 & 1 & 1 & 3 & X_3 \\
        0 & 0 & 0 & 1 & 1 & 1 & 1 & 0 & 0 & 1 & 1 & 0 & 0 & 1 & 1 & 1 & 1 & 0 & 1 & 0 & 1 & 1 & 1 & 3 & 1 & X_4
    \end{pNiceMatrix}.
\end{equation}

With $\textbf{A}_{(BZ-\Gamma)}$ and $\mathbf{v}^T_{(BZ-\Gamma)}$ in hand, we can now solve the Diophantine equation following one of the approaches described in the main text. 
The solution will be a vector of length equal to the number of EBRs in the SG (25 in this particular case), indicating the sum of EBRs that will describe the symmetry content of the band structure. In our example, the minimal solution will be

\begin{align}
    \mathbf{n}^T & = A_{2u}@4b + A_{2u}@4c - A_1@2a \\
    & = (-1,0,0,0,0,0,0,0,1,0,0,0,0,0,1,0,0,0,0,0,0,0,0,0,0),
\end{align}
where each entry in the vector is associated with an EBR, following the order indicated by the columns matrix $\textbf{A}_{(BZ-\Gamma)}$. In this case

\begin{align}
    \{A_1@2a, A_2@2a, E@2a, T_1@2a, T_2@2a, A_{1g}@4b, A_{1u}@4b, A_{2g}@4b, A_{2u}@4b, E_{g}@4b, E_u@4b, \\
    A_{1g}@4c, A_{1u}@4c, A_{2g}@4c, A_{2u}@4c, E_g@4c, E_u@4c, A_1@6d, A_2@6d, B_1@6d, B_2@6d, A_1@12f, \\
    B_1@12f, B_2@12f, B_3@12f\}.
\end{align}

\subsubsection{TETB model}

Here, we give completely explicit formulas for the  Hamiltonian of the TETB model with SG $Pn\Bar{3}m$ (No. 224).
As discussed in the main text, this can be written as a time reversal symmetric piece plus a linear perturbation in the external magnetic field, $H_M(\textbf{k},\textbf{H})=H(\textbf{k})+f(\textbf{k},\textbf{H})$. 
The magnetic field independent piece of the Hamiltonian is given by

{\scriptsize
\begin{equation}    
    H(\textbf{k})=\begin{pmatrix}
        \alpha_1 & f(k_x,k_y) & f(k_x,k_z) & f(k_y,k_z) & g(k_x,k_y,k_z) & p(k_z) & p(k_y) & p(k_x) \\
        f(k_x,k_y) & \alpha_1 & f(k_y,-k_z) & f(k_x,-k_z) & p(k_z) & g(k_y,k_x,-k_z) & p(k_x) & p(k_y) \\
        f(k_x,k_z) & f(k_y,-k_z) & \alpha_1 & f(k_x,-k_y) & p(k_y) & p(k_x) & g(k_z,k_y,-k_x) & p(k_z) \\
        f(k_y,k_z) & f(k_x,-k_z) & f(k_x,-k_y) & \alpha_1 & p(k_x) & p(k_y) & p(k_z) & g(-k_y,k_x,-k_z) \\
        g(k_x,k_y,k_z) & p(k_z) & p(k_y) & p(k_x) & \alpha_2 & q(k_x,k_y) & q(k_x,k_z) & q(k_y,k_z) \\
        p(k_z) & g(k_y,k_x,-k_z) & p(k_x) & p(k_y) & q(k_x,k_y) & \alpha_2 & q(k_y,-k_z) & q(k_x,-k_z) \\
        p(k_y) & p(k_x) & g(k_z,k_y,-k_x) & p(k_z) & q(k_x,k_z) & q(k_y,-k_z) & \alpha_2 & q(k_x,-k_y) \\
        p(k_x) & p(k_y) & p(k_z) & g(-k_y,k_x,-k_z) & q(k_y,k_z) & q(k_x,-k_z) & q(k_x,-k_y) & \alpha_2 \\
    \end{pmatrix},
\end{equation}
}
where

\begin{equation}
    \begin{aligned}
        f(x_1,x_2)= & a_2\cos\left(\pi[x_1+x_2]\right)+r_2\cos\left(\pi[x_1-x_2]\right) \\
        g(x_1,x_2,x_3)= & a_3\cos\left(\pi[x_1+x_2+x_3]\right)+r_3\cos\left(\pi[x_1-x_2-x_3]\right)+2r_3\cos\left(\pi x_1\right)\cos\left(\pi[x_2-x_3]\right) \\
        p(x_1)= & a_1\cos\left(\pi x_1\right) \\
        q(x_1,x_2)= & s_2\cos\left(\pi[x_1+x_2]\right)+w_2\cos\left(\pi[x_1-x_2]\right) \\
    \end{aligned},
\end{equation}
and $\alpha_1,\alpha_2,a_2$, $r_2$, $a_3$, $r_3$, $a_1$, $s_2$ and $w_2$ are real, independent parameters. The linear perturbation in $\textbf{H}$ depends on the five real parameters $\beta_2$, $\delta_2$, $\delta_1$, $\kappa_2$ and $\epsilon_2$ and is given by

\newpage

\begin{sideways}
    \parbox{\textheight}{
    
    \vspace{0.25\textwidth}
    {\scriptsize
    \begin{equation}
        \begin{aligned}
            & f(\textbf{k},\textbf{H})= \\
            & \begin{pmatrix}
            0 & h(k_x,k_y,H_x,H_y) & h(k_z,k_x,H_z,H_x) & h(k_y,k_z,H_y,H_z) & 0 & g(k_z,H_x,H_y) & g(k_y,H_z,H_x) & g(k_x,H_y,H_z) \\
            h^*(k_x,k_y,H_x,H_y) & 0 & h(-k_y,k_z,-H_y,H_z) & h(k_x,-k_z,-H_x,H_z) & g(k_z,H_y,H_x) & 0 & g(k_x,-H_y,H_z) & g(k_y,H_z,-H_x) \\
            h^*(k_z,k_x,H_z,H_x) & h^*(-k_y,k_z,-H_y,H_z) & 0 & h(-k_x,k_y,-H_x,H_y) & g(k_y,H_x,H_z) & g(k_x,H_y,-H_z) & 0 & g(k_z,-H_x,H_y) \\
            h^*(k_y,k_z,H_y,H_z) & h^*(k_x,-k_z,-H_x,H_z) & h^*(-k_x,k_y,-H_x,H_y) & 0 & g(k_x,H_z,H_y) & g(k_y,-H_x,H_z) & g(k_z,H_x,-H_y) & 0 \\
            0 & g^*(k_z,H_y,H_x) & g^*(k_y,H_x,H_z) & g^*(k_x,H_z,H_y) & 0 & p(k_x,k_y,H_x,H_y) & p(k_z,k_x,H_z,H_x) & p(k_y,k_z,H_y,H_z) \\
            g^*(k_z,H_x,H_y) & 0 & g^*(k_x,H_y,-H_z) & g^*(k_y,-H_x,H_z) & p^*(k_x,k_y,H_x,H_y) & 0 & p(-k_y,k_z,-H_y,H_z) & p(k_x,-k_z,-H_x,H_z) \\
            g^*(k_y,H_z,H_y) & g^*(k_x,-H_y,H_z) & 0 & g^*(k_z,H_x,-H_y) & p^*(k_z,k_x,H_z,H_x) & p^*(-k_y,k_z,-H_y,H_z) & 0 & p(-k_x,k_y,-H_x,H_y) \\
            g^*(k_x,H_y,H_z) & g^*(k_y,H_z,-H_x) & g^*(k_z,-H_x,H_y) & 0 & p^*(k_y,k_z,H_y,H_z) & p^*(k_x,-k_z,-H_x,H_z) & p^*(-k_x,k_y,-H_x,H_y) & 0 \\
        \end{pmatrix}
        \end{aligned},
    \end{equation}
    }
    where
    
    \begin{equation}
        \begin{aligned}
            h(x_1,x_2,H_1,H_2)= & i(H_1-H_2)\left\{\beta_2\cos\left(\pi[x_1-x_2]\right)+\delta_2\cos\left(\pi[x_1+x_2]\right)\right\} \\
            g(x_1,H_1,H_2)= & i\delta_1(H_1-H_2)\cos\left(\pi x_1\right) \\
            p(x_1,x_2,H_1,H_2)= & i(H_1-H_2)\left\{\kappa_2\cos\left(\pi[x_1-x_2]\right)+\epsilon_2\cos\left(\pi[x_1+x_2]\right)\right\} \\
        \end{aligned},
    \end{equation}
    }
\end{sideways}

\subsubsection{Consistency check using magnetic SGs}

As we mentioned in the main text, the linear perturbation that we introduced to mimic the behavior of a magnetic field must be consistent with a symmetry analysis of the system after breaking TRS because of the introduction of a magnetic field. Since TRS is broken, in order to describe the system symmetries, we will employ magnetic SGs \cite{perez2015symmetry}.

If we consider that a magnetic field is applied along the $z$-direction, the only symmetries that will be preserved can be obtained by inspection, and they will be 

\begin{itemize}
    \item The 2-fold and 4-folds rotations along the $z$-axis,
    \item The 2-folds times time-reversal symmetry on the $xy$-plane and
    \item Inversion symmetry.
\end{itemize}

This set of operations is mapped to MSG $p4_2/nn'm'$ (No. 134.477) by a shift of $1/2$ along the $y$-direction. To identify the mapping, the Bilbao Crystallographic Server (BCS) \cite{elcoro2017double} provides a helpful tool in \textit{Representations and Applications $\to$ Magnetic point and space groups $\to$ MCORREL}.

With this identification of the MSG, we can subduce the symmetry vector of our non-magnetic band structure ($\mathbf{v}^T_{NM}$) from the non-magnetic supergroup to the magnetic subgroup to obtain the magnetic symmetry vector ($\mathbf{v}^T_{M}$). 

In our example, $\mathbf{v}^T_{NM}$ is (Eq. (12) in the main text)

\begin{equation}
    \begin{aligned}
        \mathbf{v}^T_{NM} & = A_{2u}@4b + A_{2u}@4c - A_{1}@2a \\
        & = [-\Gamma_1^+(1) + \Gamma_2^-(1) + 2\Gamma_4^-(3),R_4^-(3) + R_5^+(3), M_1(2) + 2M_4(2), X_1(2) + X_3(2) + X_4(2)],
    \end{aligned}
\end{equation}
where we have represented the irrep alongside its multiplicity.

Doing the subduction into the MSG, we obtain the following magnetic symmetry vector 

\begin{equation}
    \begin{aligned} \label{eq:mag_subducction}
        \mathbf{v}^T_M & = A_u@4e + A_u@4f - A@2a \\
        & = [-\Gamma_1^+(1) + 2\Gamma_1^-(1) + \Gamma_2^-(1) + 2\Gamma_3^-(1) + 2\Gamma_4^-(1), A_1^-(1) + A_2^+(1) + A_3^+(1) \\
        & \quad + A_3^-(1) + A_4^+(1) + A_4^-(1), 3R_1(2), M_1(2) + 2M_2(2), Z_1(2) + 2Z_2(2), 3X_1(2)].
    \end{aligned}
\end{equation}

Now, we can verify that this result is consistent with the magnetic band structure obtained from Maxwell's eigensolver (MPB). Similarly, it must also be equivalent to that obtained following our perturbation approach and particularizing the magnetic field to one pointing in the $z$-direction. 

Computing the irreps from the symmetry eigenvalues of the eigenvectors obtained from MPB, we find the following symmetry vector

\begin{equation}
    \begin{aligned}
        \mathbf{v}^T_{(BZ-\Gamma),\text{MPB}} & = [A_1^-(1) + A_2^+(1) + A_3^+(1) + A_3^-(1) + A_4^+(1) + A_4^-(1), \\
        & \quad \; 3R_1(2), M_1(2) + 2M_2(2), Z_1(2) + 2Z_2(2), 3X_1(2)],
    \end{aligned}
\end{equation}

Additionally, if we include the symmetry content at the $\Gamma$ point, by carefully analyzing the ill-defined value at zero-frequency proposed in Ref. \cite{christensen2022location}, the complete symmetry vector takes the following form: 

\begin{equation}
    \begin{aligned}
        \mathbf{v}^T_{\text{MPB}} & = [-\Gamma_1^+(1) + 2\Gamma_1^-(1) + \Gamma_2^-(1) + 2\Gamma_3^-(1) + 2\Gamma_4^-(1), A_1^-(1) + A_2^+(1) + A_3^+(1) \\
        & \quad + A_3^-(1) + A_4^+(1) + A_4^-(1), 3R_1(2), M_1(2) + 2M_2(2), Z_1(2) + 2Z_2(2), 3X_1(2)].
    \end{aligned}
\end{equation}

This symmetry vector is identical to the one we obtain following the perturbation approach and particularizing the magnetic field to one pointing in the $z$-direction (see Figure 2 in the main text), and it is in perfect agreement with the one we derived subducing $\mathbf{v}^T_{NM}$ into the MSG (Eq. \eqref{eq:mag_subducction}).

This analysis proves that we did not make any mistake constructing the model for this particular example, and it is a good addition as a consistency check.

\subsection{TETB Hamiltonian for SG221} \label{sec:NM_221}

\subsubsection{Solution to the Diophantine equations} \label{subsec:rsi221}

Here we apply the RSI analysis to SG $Pm\bar{3}m$ (No. 221).
Similar to the discussion on SG $Pn\bar{3}m$ (No. 224), we only focus on the RSIs defined for maximal WPs.
The relevant WPs are $1a: (0,0,0)$ with site-symmetry group $G_{1a}=m\bar{3}m$, $1b: (1/2,1/2,1/2)$ with $G_{1b}=m\bar{3}m$, $3c: (0,1/2,1/2)$ with $G_{3c}=4/mmm$, and $3d: (1/2,0,0)$ with $G_{3d}=4/mmm$.
For these maximal WPs, the RSIs are defined by

\begin{align}
    \delta_{\mathbf{q},1} =& -m(A_{1g}) + m(A_{1u}) - m(T_{1g}) + m(T_{1u}), \nonumber \\
    \delta_{\mathbf{q},2} =& -m(A_{1g}) + m(A_{1u}) - m(A_{2g}) + m(A_{2u}) + m(E_g) - m(E_u), \nonumber \\
    \delta_{\mathbf{q},3} =& -m(A_{2g}) + m(A_{2u}) - m(T_{2g}) + m(T_{2u}), \nonumber \\
    \delta_{\mathbf{q}',1} =& m(A_{1g}) - m(A_{1u}) - m(B_{1g}) + m(B_{1u}) + m(A_{2g}) - m(A_{2u}) -m(B_{2g}) + m(B_{2u}), \nonumber \\
    \delta_{\mathbf{q}',2} =& -m(B_{1g}) + m(B_{1u}) - m(B_{2g}) + m(B_{2u}) - m(E_g) + m(E_u),
\end{align}
where $\mathbf{q}=1a,1b$ and $\mathbf{q}'=3c,3d$.

\begin{figure*}
    \centering
    \includegraphics[width=\textwidth]{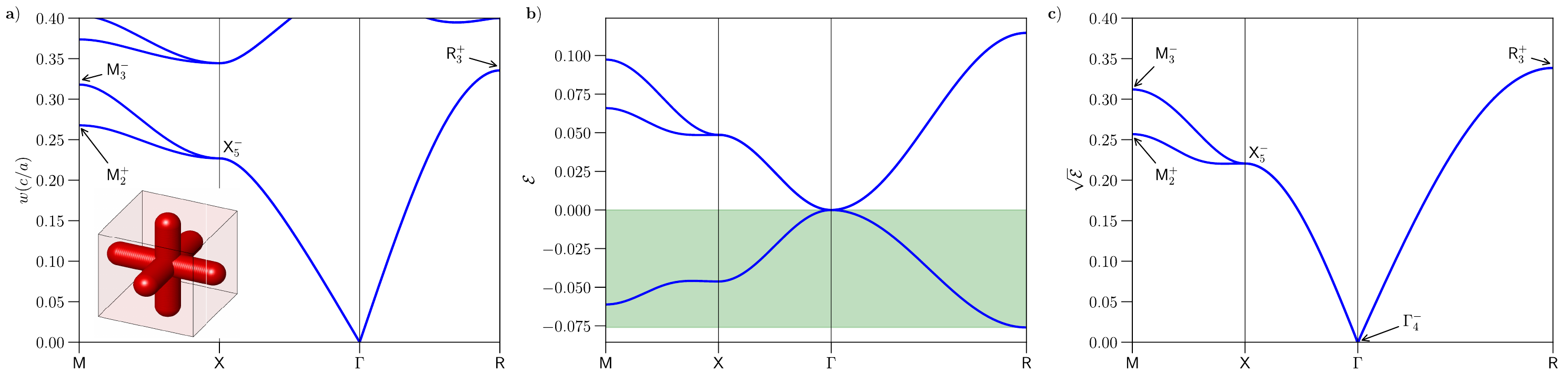}
    \caption{Band structures obtained from MPB and TETB model for a crystal with SG $Pm\Bar{3}m$ (No. 221).
    \textbf{a)} Frequency bands obtained from the crystal.
    \textbf{b)} Band structure obtained from the TB model.
    The bands enclosed by the green-shaded region belong to the additional modes included to regularize the symmetry content at $\Gamma$.
    \textbf{c)} Band structure obtained from the TB model after applying the mapping to the electromagnetic-wave equation frequency and imposing the transversality constraint.}
    \label{fig:panel221}
\end{figure*}

We are interested in building  a TETB model for the photonic system in Fig. \ref{fig:panel221}\textcolor{red}{a} based on the symmetry vector

\begin{gather}\label{eq:symv2}
    \mathbf{v}^T = [(\blacksquare)^{2T},R_3^+,M_2^+ + M_3^-,X_5^-],
\end{gather}
with $(\blacksquare)^{2T} = \Gamma_4^- - \Gamma_1^+$, for which the following indices are obtained 
\begin{align}
    (\delta_{a,1},\delta_{a,2},\delta_{a,3},\delta_{b,1},\delta_{b,2},\delta_{b,3},\delta_{c,1},\delta_{c,2},\delta_{d,1},\delta_{d,2}) = (1,1,0,0,0,0,0,0,-1,0).
\end{align}
Thus, only the WPs $1a$ and $3d$ have nonzero RSIs.
Noting that ${\rm dim}(\mathbf{v}^T)=2$, equal to the filling or number of bands, and taking   $(\blacksquare)^{2T}=-\Gamma_1^+ + \Gamma_4^-$, we find $\mathbf{n}^T = A_{2u}@{3d}  -  A_{1g}@{1a}$.
One can further confirm that this EBR decomposition is consistent with the full RSI indices. The same result is obtained by the optimization and enumeration methods.

\subsubsection{TETB Model} \label{subsec:tb221}

Unlike our previous example, a look at Fig. \ref{fig:panel221}\textcolor{red}{a} shows the existence of a global (fundamental) gap separating the two
lowest frequency bands from the rest. This agrees with Ref. \cite{christensen2022location}, which gives $\mu_1^T=2$ with two possible symmetry vectors, one of which matches Eq. \eqref{eq:symv2}. As all the symmetry indicators for $Pm\bar{3}m$ (No. 221) are trivial, there is no obstacle to building a TETB model based on the EBR $A_{2g}@3d$. Thus, we can develop our TETB model by placing pseudo-orbitals transforming as $A_{2g}$ at the WP $3d$: $(1/2,0,0)$ with site-symmetry group $4/mmm$. They will be denoted by
$d_i(\textbf{r})$, where $i=1,2,3$ labels the three positions in WP $3d$.
As explained in the main text, the Hamiltonian has to satisfy the set of constraints
\begin{equation}\label{eq:generalTransform}
    D(g)H(\textbf{k})D(g)^{-1}=H(g\textbf{k})
\end{equation}
and
\begin{equation}\label{eq:time}
  H(-\textbf{k})=H(\textbf{k})^*.
\end{equation}
where $g$ runs over the generators of SG and $D(g)$ stands for the matrix of the representation defined on $\{d_i(\textbf{r})\}$. According to the BCS, a possible set of generators is given by $\{1|0\}$, $\{2_{001}|0\}$, $\{2_{010}|0\}$, $\{3_{111}^+|0\}$, $\{2_{110}|0\}$ and $\{\bar 1|0\}$. 

To build $D(g)$, we begin by choosing a reference WP, e.g. position $\textbf{q}_1$, and denote by $G_1$ its site-symmetry group. Here, we have $G_1 \cong D_3$, where $ \cong$ denotes an isomorphism.
Let us denote by $G_1$ the site-symmetry group of the first position $\textbf{q}_1$ in WP $3d$: $(1/2,0,0)$.
For each position $\textbf{q}_\alpha$ in WP $3d$ there must exist a SG element $g_\alpha$ such that $\textbf{q}_\alpha=g_\alpha \textbf{q}_1$.
Then, given a representation $\rho_1$ of $G_1$, which can be found in the BCS, we can construct the induced representation $\rho$ of the complete SG by
\begin{equation}
    [\rho(g)]_{(i,\alpha,\textbf{t}),(j,\beta,\textbf{t}')}=[\Tilde{\rho}(g_\alpha^{-1}\{1|\textbf{t}\}g\{1|\textbf{t}'\}^{-1}g_\beta)]_{ij},
\end{equation}
where $g$ is any transformation of the SG and
\begin{equation}
    [\Tilde{\rho}(h)]_{ij}=\left\{\begin{matrix}
        [\rho_1(h)]_{ij} & \text{if $h\in G_1$} \\
        0 & \text{else}
    \end{matrix}\right.
\end{equation}
The resulting action of the generators of SG $Pm\Bar{3}m$ (No. 221) on the basis elements are summarized in Table \ref{tab:orbTransf}.

\begin{table}
    \centering
    \begin{tabular}{c|cccccc}
         \multicolumn{1}{c}{} & $\{1|0\}$ & $\{2_{001}|0\}$ & $\{2_{010}|0\}$ & $\{3_{111}^+|0\}$ & $\{2_{110}|0\}$ & $\{-1|0\}$ \\\hline\hline
        $d_{1}$ & $d_{1}$ & $-d_{1}$ & $-d_{1}$ & $d_{2}$ & $d_{2}$ & $d_{1}$ \\
        $d_{2}$ & $d_{2}$ & $-d_{2}$ & $d_{2}$ & $d_{3}$ & $d_{1}$ & $d_{2}$ \\
        $d_{3}$ & $d_{3}$ & $d_{3}$ & $-d_{3}$ & $d_{1}$ & $-d_{3}$ & $d_{3}$
    \end{tabular}
    \caption{
 Transformation of the pseudo-orbitals under the generators of SG $Pm\Bar{3}m$ (No. 221).}
    \label{tab:orbTransf}
\end{table}

Finally, one can write the action of the generators of the SG on the pseudo-orbitals in momentum space as 
\begin{equation}\label{eq:repreK}
    \rho(g)d_i(\textbf{k},\textbf{r})=e^{-i(R\textbf{k})\cdot\textbf{v}}\sum_{j}[\rho_1(h)]_{ji}d_j(R\textbf{k},\textbf{r}),
\end{equation}
where $g=\{R|\textbf{v}\}$ with $R$ a point group element and $\textbf{v}$ a translation.
Note that the expression in Eq. \eqref{eq:repreK} will depend on how the Fourier transform is defined.
Here, we have chosen $d_i(\textbf{k},\textbf{r})=\sum_\textbf{t}e^{i\textbf{k}\cdot(\textbf{t}+\textbf{r}_i)}d_i(\textbf{r}-\textbf{t})$, where $\textbf{r}_i$ is the orbital position. On this basis, the generators are represented by the matrices

\begin{equation}\label{eq:221transfor}
    \begin{aligned}
        D(\{1|0\})&=\begin{pmatrix}
        1 & 0 & 0 \\
        0 & 1 & 0 \\
        0 & 0 & 1 \\
    \end{pmatrix}\qquad D(\{2_{001}|0\})=\begin{pmatrix}
        -1 & 0 & 0 \\
        0 & -1 & 0 \\
        0 & 0 & 1 \\
    \end{pmatrix} \\[0.2cm]
    D(\{3_{111}^+|0\})&=\begin{pmatrix}
        0 & 0 & 1 \\
        1 & 0 & 0 \\
        0 & 1 & 0 \\
    \end{pmatrix}\qquad D(\{2_{010}|0\})=\begin{pmatrix}
        -1 & 0 & 0 \\
        0 & 1 & 0 \\
        0 & 0 & -1 \\
    \end{pmatrix} \\[0.2cm]
    D(\{2_{110}|0\})&=\begin{pmatrix}
        0 & 1 & 0 \\
        1 & 0 & 0 \\
        0 & 0 & -1 \\
    \end{pmatrix}\qquad D(\{-1|0\})=\begin{pmatrix}
        -1 & 0 & 0 \\
        0 & -1 & 0 \\
        0 & 0 & -1 \\
    \end{pmatrix}.
    \end{aligned}
\end{equation}

Then, solving the constraints in Eqs. \eqref{eq:generalTransform} and \eqref{eq:time} yields 
\begin{equation} \label{eq:221H}
        H(\textbf{k})=
            \begin{pmatrix}
                f(k_x,k_y,k_z) & g(k_x,k_y) & g(k_x,k_z) \\
                g(k_x,k_y) & f(k_y,k_x,k_z) & g(k_y,k_z) \\
                g(k_x,k_z) & g(k_y,k_z) & f(k_z,k_x,k_y)
            \end{pmatrix},
\end{equation}
where
\begin{equation}
    \begin{aligned}
         f(x_1,x_2,x_3)&=\alpha+2a_2\cos\left(2\pi x_1\right)+2b_2\left\{\cos\left(2\pi x_2\right)+\cos\left(2\pi x_3\right)\right\} \\[0.2cm]
         g(x_1,x_2)&=4a_1\sin\left(\pi x_1\right)\sin\left(\pi x_2\right).
    \end{aligned}
\end{equation}
This Hamiltonian depends on four real parameters ($\alpha$, $a_1$, $a_2$, $b_2$), where $\alpha$ is on-site energy,  while the remaining three parameters are first ($a_1$) and second ($a_2$, $b_2$) nearest neighbor hoppings. Imposing $H(\textbf{0})=0$ yields $\alpha = -2a_2-4b_2$.
Table \ref{tab:221} gives the values of the three independent parameters that best fit the numerical simulation.
\begin{table}
    \centering
    \begin{tabular}{cc}
        Parameters & Value \\\hline\hline
        $a_1$ & -0.01588775 \\
        $a_2$ & 0.01156738 \\
        $b_2$ & -0.01216815
    \end{tabular}
    \caption{Independent parameters of the TB Hamiltonian for SG $Pm\Bar{3}m$ (No. 221).}
    \label{tab:221}
\end{table}

Note that,  as shown in Fig.~\ref{fig:panel221}\textcolor{red}{c}, in the TETB model, the bands at the origin transform according to $\Gamma_4^-$, which is the vector representation ($V=\Gamma_4^-$) for SG221. Thus, according to our discussion in subsection \ref{subsec:role}, we can use the TETB Hamiltonian to obtain a $\mathbf{k}\cdot\mathbf{p}$ description of the long wavelength, low-frequency limit of the photonic modes. Indeed, taking the $k\to 0$ limit of $H(\textbf{k})$ gives

\begin{equation} \label{eq:221kp}
        H_{k\cdot p}(\textbf{k})= 
             4 \pi^2\begin{pmatrix}
                -a_2 k_x^2 - b_2 (k_y^2 + k_z^2) & a_1 k_x k_y & a_1 k_x k_z \\
                a_1 k_y k_x &  -a_2 k_y^2 - b_2 (k_z^2 + k_x^2) & a_1 k_y k_z \\
                a_1 k_z k_x& a_1 k_z k_y&  -a_2 k_z^2 - b_2 (k_x^2 + k_y^2)
            \end{pmatrix} +\mathcal{O}(k^3),
\end{equation}

The spectrum can be analytically computed for $\textbf{k}$ along high symmetry directions. For instance, for
$\textbf{k}=(k,0,0)$ the result is $\omega_1^2=\omega_2^2= -4\pi^2 b_2 k^2$, with eigenvectors $(0,1,0)$
and $(0,0,1)$, and $\omega_3^2= -4\pi^2 a_2 k^2$ with eigenvector $(1,0,0)$. Thus, for the values in Table \ref{tab:221}, we find two degenerate transverse modes with real frequency and one longitudinal, imaginary frequency mode.

\bibliography{references.bib}